\documentclass{aa}  

\usepackage{enumitem}
\usepackage{graphicx}
\usepackage{natbib}
\usepackage{comment}
\usepackage{txfonts}
\usepackage[colorlinks = true,
            linkcolor = blue,
            urlcolor  = blue,
            citecolor = blue,
            anchorcolor = blue]{hyperref}
\begin{document}

  \title{Probing the Formation of Megaparsec-scale Giant Radio Galaxies (I): Dynamical Insights from MHD Simulations}
  \titlerunning{MHD Modelling of Giant Radio Galaxies}

   \author{Gourab Giri,\inst{1,2}
   Joydeep Bagchi,\inst{3}
   Kshitij Thorat,\inst{1}
   Roger P. Deane,\inst{1,4}
   Jacinta Delhaize,\inst{5}
   D. J. Saikia\inst{6}
          }

   \institute{Department of Physics, University of Pretoria, Private Bag X20, Hatfield 0028, South Africa
   \and
   South African Radio Astronomy Observatory, 2 Fir St, Black River Park, Observatory 7925, South Africa
   \and
    Department of Physics and Electronics, Christ University, Hosur Road, Bangalore 560029, India
    \and
    Wits Centre for Astrophysics, School of Physics, University of the Witwatersrand, 1 Jan Smuts Avenue, Johannesburg, 2000, South Africa
    \and
    Department of Astronomy, University of Cape Town, Private Bag X3, Rondebosch 7701, South Africa
    \and
    Inter-University Centre for Astronomy and Astrophysics, Post Bag 4, Pune 411007, India \\
    \email{gourab.giri@up.ac.za}
             }

   \authorrunning{G. Giri et al.}
 
  \abstract
   {Giant radio galaxies (GRGs), constituting a relatively small fraction of the extended-jetted population, form in a wide range of jet and environment configurations. This observed diversity complicates the identification of the growth factors that facilitate their attainment of megaparsec-scales.}
   {This study aims to numerically investigate the hypothesized formation mechanisms of GRGs extending $\gtrsim 1$ Mpc to assess their general applicability.}
   {We employ tri-axial ambient medium settings to generate varying levels of jet frustration and simulate jets with low and high power from different locations in the environment. This formulates five representations evolving in a relativistic magneto-hydrodynamical framework.}
   {The emergence of distinct giant phases in all five simulated scenarios suggests that GRGs may be more common than previously believed. This prediction can be verified with contemporary and forthcoming radio telescopes. We find that different combinations of jet morphology, power, and the evolutionary age of the formed structure hold the potential to elucidate different formation scenarios. In all of these cases, the lobes are over-pressured, prompting further investigation into pressure profiles when jet activity ceases, potentially distinguishing between relic and active GRGs. We observed a potential phase transition in giant radio galaxies, marked by differences in lobe expansion speed and pressure variations compared to their smaller evolutionary phases. This suggests the need for further investigation across a broader parameter space to determine if lobe evolution in GRGs fundamentally differs from smaller RGs. The axial ratio analysis reveals self-similar expansion in rapidly propagating jets, while there is a notable deviation while the jet forms wider lobes. Overall, this study emphasizes that multiple growth factors simultaneously at work can better elucidate the current-day population of GRGs, including scenarios e.g., the growth of GRGs in dense environments, GRGs extending several megaparsecs, development of GRGs in low-powered jets, and the formation of morphologies like GRG-XRGs.}
   {}


   \maketitle
%

\section{Introduction}
A limited fraction of Active Galactic Nuclei (AGN) with accreting supermassive black holes \citep[SMBHs with mass $\geq 10^{6.5} M_{\odot}$;][]{Baldi2021} at their centers exhibit the ejection of highly collimated plasma flows, with bulk Lorentz factors reaching 7 or higher \citep{Kellermann1989,Merloni2007,Padovani2011,Saikia2016}. Therefore, considerable focus has been placed on understanding the extent to which such high-speed jets can propagate, given the significant jet hindrance caused by the denser ambient medium through which the jets travel. The studies have revealed a wide range of jet lengths \citep[cf.][for a recent review]{Saikia2022} with some ending in the sub-kiloparsec scales \citep[Compact Radio Sources (CRSs);][]{Baldi2023}, some extending to galactic scales \citep[Galaxy-scale jets (GSJs);][]{Webster2021,ODea2021}, some propagating to hundreds of kiloparsecs \citep[extended kpc-scale jets;][]{Hardcastle2020}, and a small fraction evolving on megaparsec scales  \citep[Giant Radio Galaxies (GRGs);][]{Dabhade2023}. While determining the availability of such sources, large surveys have reported that the longer the jet extends, the rarer these sources become \citep{Hardcastle2016,Ishwara-Chandra2020,Gurkan2022}. Being rare, these extended radio sources attract significant attention because of their extreme physical conditions, encouraging modeling efforts on various aspects such as the diverse ambient environments they propagate through and different jet propagation models that help them cover such distances. 
Important aspect of studying extended radio jets is also that they can
provide energetic feedback to the surrounding medium \citep{Fabian2012,McNamara2012,Heckman2023}. For an example, if kinetic energy of the outflow is efficiently deposited in the environment, it can significantly affect the star formation rate \citep{Nesvabda2021,Mulard2023,Roy2024}.

This paper focuses on understanding giant radio galaxies, which are recognised as sources with jetted structures extending to at least 700 kpc. Initially, the length cutoff for GRGs was fixed at $\geq 1$ Mpc when the value of the Hubble constant was accepted to be $H_0 = 50$ $\rm km\ s^{-1}\ Mpc^{-1}$ \citep{Willis1974}. Many of the recent authors have tried to reset the length cutoff for GRGs again as $\geq 1$ Mpc with the present day estimation of Hubble constant as $H_0 = 70$ $\rm km\ s^{-1}\ Mpc^{-1}$ \citep{Andernach2021,Hardcastle2016,Bruggen2021}. Regardless of the length threshold, the origin of such extended sources appears to be ambiguous, supported by the fact that they constitute a relatively small fraction of radio-loud AGNs \citep{Dabhade2020a,Dabhade2020b}, with sources having linear extents exceeding 3 Mpc being a handful \citep{Andernach2021}. Giant radio sources reaching a total extent of 5 Mpc are extremely rare \citep{Willis1974,Machalski2008,Oei2022}, raising the question of whether there is a physical size limit to how far a jet can propagate. The recent discovery of an AGN jet spanning approximately 7 Mpc has introduced additional challenges to this topic \citep{Oei2024b}.

With advancements in the sensitivity and resolution of radio telescopes, the discovery of GRGs has increased, yet they still represent only a relatively small fraction of the extended jetted population \citep{Andernach2021,Bhukta2022,Oei2023,Mostert2024}. With these advancements in source statistics, the hypothesized formation mechanisms generating GRGs have been revised to incorporate new data and insights. Currently, two models are being considered to explain the origin of GRGs: one attributes their growth to propagation in a less-dense ambient medium, while the other links their formation to complex nuclear activity \citep[for a review, see,][]{Dabhade2023}.

The former model suggests that GRG growth occurs in the low-density Warm-Hot Intergalactic Medium (WHIM), typically found in the galaxy filaments of the Cosmic Web \citep{Dave2001,Oei2024}, or that GRGs evolve near the edge of a galaxy group or cluster medium. Significant support for this model has been found in statistical studies where available cluster and filament catalogs have been cross-matched or where polarization measures have been conducted \citep{Stuardi2020,Oei2023,Simonte2024}, as well as in focused source studies that use the galaxy distribution around the hosts of such sources to infer their large-scale environments \citep{Subrahmanyan2008,Safouris2009,Pirya2012,Malarecki2015,Oei2022}. Additional evidence comes from the observation that GRGs also exhibit off-axis lobe distortions, linking to the presence of an underlying asymmetric ambient medium  \citep{Subrahmanyan1996,Malarecki2013}. Furthermore, the lengths of the two lobes are typically dissimilar and the ratio of bigger to smaller lobe lengths are marginally greater than those of their Smaller Radio Galaxy counterparts (SRGs; $< 0.7$ Mpc), which cannot be fully explained by orientation effects alone \citep{Saripalli1994,Ishwara-Chandra1999,Schoenmakers2000}.

This model is however not free from caveats, as studies have found that $\sim 10$ percent of GRGs are evolving within a group or cluster medium, with many of them acting as the brightest cluster galaxies \citep{Komberg2009,Dabhade2020b,Andernach2021,Sankhyayan2024}. The studies by \citet{Lan2021} and \citet{Simonte2022} found no preference for ambient environment factors, such as the location of the host galaxy with respect to the surrounding large-scale structure or the properties of nearby galaxies, that would favor the growth of GRGs over SRGs. Moreover, one needs to also confirm that the galaxy distribution used as a proxy for the gas distribution is in general true; an exception to this may be seen in \citet{Palma2000}.

The latter model, which involves activity near the host AGN, includes possibilities such as a high-power jet ejection, a subsequently longer jet ejection time, or restarting activity that helps the lobes grow at a considerable rate. The possibility that recurrent activity of AGN contributes to the growth of GRGs seems unlikely to be a common occurrence, given that the availability of such cases among GRGs is low \citep[$\sim 5$ percent,][]{Dabhade2020a,Dabhade2020b}. The fraction of extended jets exhibiting restarting activity is low in general \citep[13 - 15\%;][]{Jurlin2020}. A longer jet evolution time appears to be an important aspect and has been considered in several studies, which have indicated evidence of a greater jet evolution age for GRGs compared to their smaller counterparts \citep[e.g.,][]{Jamrozy2008,Komberg2009,Machalski2011,Dabhade2023}. However, several studies have also estimated relatively younger (spectral) ages for GRGs \citep{Mack1998,Konar2004}, comparable to SRGs \citep{Harwood2017,Turner2018}. 

The distribution of radio power (and thus jet power) of GRGs indicates that they predominantly lie above the classical \citet{Fanaroff1974} I/II boundary \citep[][]{Dabhade2020b,Andernach2021,Simonte2024}. This is often cited as the reason why the jets can overcome jet frustration (hindrance to jet motion) imposed by the ambient medium. However, if this is the case, it must also be explained why only $\sim 10$ percent of FR II radio jets evolve into GRGs \citep{Komberg2009}. Despite the lower occurrence of FR I GRGs, they also appear to overcome the opposing force of the ambient environment to the jet flow to reach such distances. Notably, majority of FR I sources reside in dynamically more evolved galaxy clusters or in rich galaxy groups \citep{Zirbel1997,Mingo2019}. The discovery of GRGs exceeding 1 Mpc in length originating from spiral or disk-type host galaxies has further complicated efforts to disentangle the connection between GRG origin and central AGN activity \citep{Hota2011,Bagchi2014,Clarke2017,Oei2023a}. Clues from host AGN black hole masses have also been inconclusive, as an extended GRG of $\sim 5$ Mpc size (`Alcyoneus') hosts a black hole of $\sim 4 \times 10^8 M_{\odot}$, while a much more massive black hole of mass $\sim 5 \times 10^{10} M_{\odot}$ (`J103129.54+502959.1') produces a jet with a length of 0.81 Mpc \citep{Oei2022,Oei2023}. As also reported by \citet{Kuzmicz2012}, the median values in black hole masses in giant radio quasars (GRQs) are not considerably different from that of smaller radio quasars. On the other hand, when compared to giant radio galaxies, GRQs are reported to have  higher black hole mass and higher Eddington ratio \citep{Mahato2022}.

A numerical approach to test these models and their caveats has thus become essential, enabling examination of jet evolution phases across a broad spectrum of ambient configurations and jet powers. This is the primary focus of this work, where we aim to determine whether giant radio galaxies are similar to their smaller counterparts or if they represent a distinct class of sources. By formulating different jet-ambient medium settings, we aim to explore the (thermo)dynamical states of the evolving intertwined structures. Given the length scales involved for GRGs, the simulations are computationally intensive, and to our knowledge, this work is the first of its kind. We intend to simulate sources with lengths extending to 1 Mpc and beyond, as it is expected that the larger the source, the more prominently the growth factors contributing to its expansion will appear \citep{Oei2022}.

This paper is structured as follows: In Section~\ref{Sec:Numerical setup}, we discuss the formulation of different simulation settings to test GRG formation models. Section~\ref{Sec:GRG properties and Observational relevance} elaborates on the results obtained from the simulations, including the morphological distinctions of the grown structures (Section~\ref{Subsec:Morphological Distinction}), their temporal evolution (Section~\ref{Subsec:Temporal signatures}), and dynamical characteristics (Section~\ref{Subsec:Dynamical characteristics}), as well as discusses their relevance to observations. We then summarize our work in Section~\ref{Sec:Summary}.

\section{Numerical setup} \label{Sec:Numerical setup}
Considering the crucial importance of modeling magnetic fields in the radio lobes of giant radio galaxies \citep{Jamrozy2008,Machalski2009}, as well as the presence of relativistic effects in the jets of GRGs \citep{Saripalli1997,Giovannini1999,Hocuk2010}, we formulated our setup using a relativistic magneto-hydrodynamical (RMHD) framework with the PLUTO astrophysical code \citep{Mignone2007}. The RMHD equations to be solved are defined as follows,
\begin{equation}
\begin{split} \label{eq:1}
    \partial_{\kappa} (\rho u^{\kappa}) = 0\\
    \partial_{\kappa} (w u^{\kappa} u^{\delta} - b^{\kappa} b^{\delta} +p g^{\kappa \delta}) = 0\\
    \partial_{\kappa} (u^{\kappa} b^{\delta} - u^{\delta}b^{\kappa}) = 0
\end{split}
\end{equation}
where $\kappa$, $\delta$ $= (0,1,2,3)$ \citep{Mignone2006,Mignone2009}. The symbols in Eq.~\ref{eq:1} represent: $\rho$ as the rest mass density, $u^{\kappa}$ as the four velocity ($\Gamma(1, \textbf{v})$: $\textbf{v}$ is the three velocity represented through $\Gamma$, the jet Lorentz factor) and $b^{\kappa}$ as the covariant magnetic field prescribed through the lab field \textbf{B} as ($b^0 , \textbf{B}/\Gamma + b^0 \textbf{v}$). Total enthalpy and total pressure of the system are defined as $w$ and $p$, respectively. The metric $g^{\kappa \delta}$ takes the form diag(-1, 1, 1, 1).

\subsection{Configurations of the ambient medium} \label{Subsec:Configurations of the ambient medium}

Substantial work aimed at understanding the ambient medium in which GRGs thrive indicates that these giant radio jets prefer to reside in cosmic filaments, smaller groups, and low-mass clusters, the density of which can be modeled using the general King’s beta-profile \citep{Machalski2009,Oei2023a}. This tri-axial density profile \citep{Cavaliere1976} can be expressed (in Cartesian system) as, 
\begin{equation} \label{eq:2}
    \rho (x',\, y',\, z) = \rho_{0} \left(1+\frac{(x'-x_0)^{2}}{a^{2}}+\frac{(y'-y_0)^{2}}{b^{2}}+\frac{(z-z_0)^{2}}{c^{2}}\right)^{-3\beta/2}
\end{equation}

\noindent Eq.~\ref{eq:2} is characterized by $\rho_{0}$, which indicates the core density of the medium, and $\beta$, which dictates the density gradient. By setting $\rho_{0}$ to 0.001 amu/cc ($\sim 200$ times the critical cosmic density of the local universe) and $\beta$ to 0.55, the particle density ranges from $10^{-3}\, {\rm cm^{-3}}$ at the core to $10^{-6}\, {\rm cm^{-3}}$ at the outskirts ($\sim 700$ kpc), consistent with typical environmental density estimated around GRG lobes \citep{Mack1998,Stuardi2020}. Given such a profile, the total baryonic mass evaluated within one megaparsec is $\sim 3.3 \, \times \, 10^{12} M_{\odot}$, and
including the dark-matter the total mass approaches  $1.9 \, \times \, 10^{13}M_{\odot}$ with a virial radius \citep{Gaspari2020} of 747 kpc,
representing a poor galaxy group or WHIM \citep{Sun2009,Oei2023a}. 

To break any symmetry that may arise while the jet propagates along the symmetry axes of the 3D configuration, we rotated the ambient medium by $10^{\circ}$ in the $x-y$ plane, keeping the $z$-axis as the rotation axis. Therefore, $x'$ and $y'$ in Eq.~\ref{eq:1} are defined as follows:
\begin{equation}\label{eq:3}
    x' = x\,{\rm{cos}}10^{\circ} - y\,{\rm{sin}}10^{\circ},\ \ y' = x\,{\rm{sin}}10^{\circ} + y\,{\rm{cos}}10^{\circ}
\end{equation}

The values of $a$, $b$, and $c$ control the tri-axial spheroid shape of the medium and represent the effective core radii, whereas $(x_0,\, y_0,\, z_0)$ represents the center of the ambient medium. To examine jet growth across a range of ambient conditions, we selected three different configurations (given that the jet always flows along the $-x$ direction) by varying these values as,
\begin{enumerate}[leftmargin=*,labelindent=\parindent]
\item Jet flow along the minor axis (`min'):\\
values of $b = 66$ kpc and $a,\,c = 33$ kpc to reduce jet frustration,
\item Jet flow along the major axis (`maj'):\\
values of $a = 66$ kpc and $b,\,c = 33$ kpc to increase jet frustration,
\item Jet flow along the edges (`edge'):\\
values of $a,\,b,\,c = 66$ kpc to create a spherical environment, then initiate jet propagation from the edge (600 kpc above host's center).
\end{enumerate}

\begin{figure*}
\centering
\includegraphics[width=2\columnwidth]{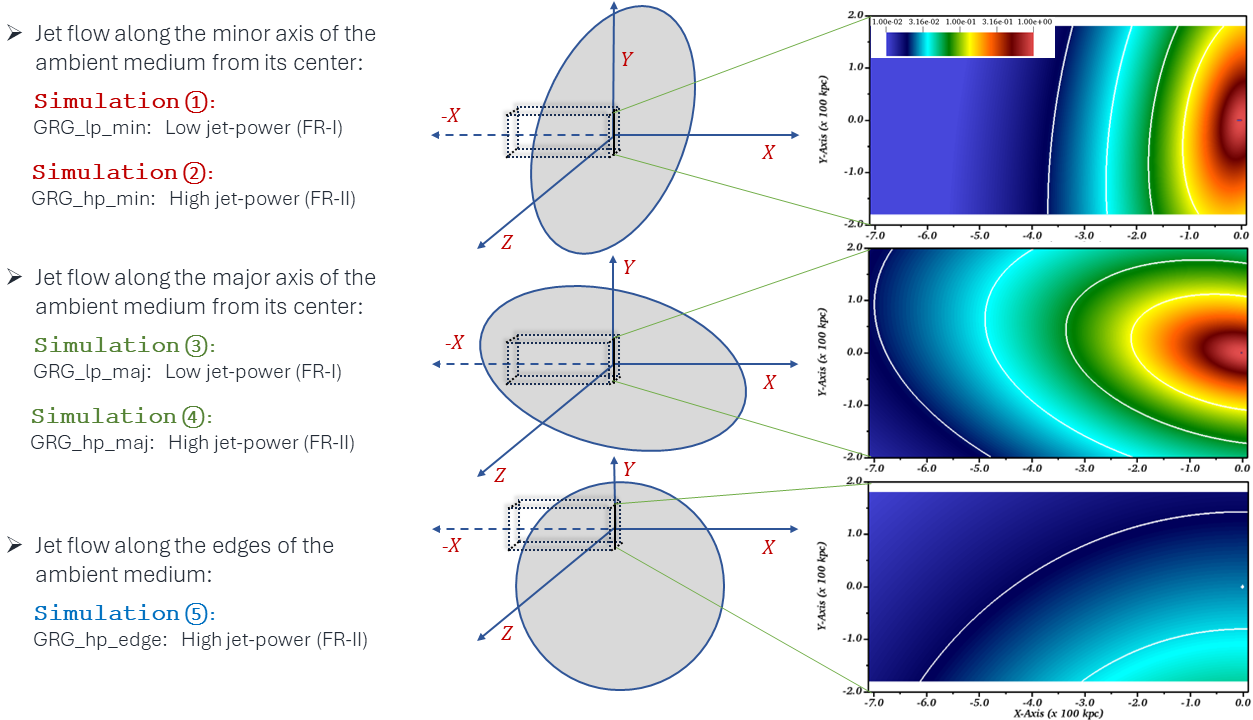}
\caption{A concise overview of the simulations conducted in this work. \textit{Left column:} a brief description of the jet-ambient medium configurations used to explore a broad spectrum of jet-environment parameters. \textit{Middle column:} the ambient configuration in relation to the Cartesian axes, with a 3D box representing the schematic simulation domain. \textit{Right column:} an $x-y$ slice of density (at $z = 0$) illustrating the initial variation of ($\rho/\rho_0$) as a color map, with overplotted pressure ($P/P_0$) contours at values of $[4.5,\, 2.2,\, 1.2,\, 0.67]\times 10^{-7}$, outwards. Jet injection zones are marked with black and white dots in the density slices. The values of $\rho_0$ and $P_0$ are 0.001 amu/cc and $1.5\times10^{-6}\, {\rm dyn/cm^2}$ respectively, and lengths are represented in units of 100 kpc. To break the symmetry with respect to the jet flow, the ambient medium has been slightly rotated.}
\label{Fig:Setup} 
\end{figure*}

\noindent The aforementioned configurations are illustrated in Fig.~\ref{Fig:Setup} through a schematic diagram and 2D slices of the initial ambient medium setup. Such choices of core radii, $\beta$, and $\rho_0$ for a galaxy group or cluster medium are typically adopted in analytical formulations of these environments \citep{Hardcastle2018,Musoke2020,Giri2023}. The tri-axial nature of the ambient medium is also not uncommon and has been observed in statistical X-ray studies of (unrelaxed) cluster sources \citep[e.g.,][]{Hodges-KLuck2010,Bruggen2021}, as well as in many individual X-ray studies of groups and cluster radio sources (as inferred through preliminary assessment) \citep{Hodges-Kluck2011,Bogdan2014,Randall2015,Ramatsoku2020,Pandge2022}. Additionally, the 3-dimensional shape and dynamics of dark matter halos around galaxies and galaxy groups or clusters is a frontier topic of ongoing research, but  numerical simulations \citep[][]{Chua2019,Emami2021,Baptista2023} and limited observations \citep[e.g.,][]{Buote2002} suggest that they are not  perfectly  spherical  but  may assume a triaxial spheroidal geometry,  with long axis preferably oriented parallel to  longer axis of  the  central  galaxy or group.  This is consequence of the  fact that the gravitational force that shapes dark matter halos are not perfectly  symmetrical, which governs their 3-dimensional  equilibrium figures in a rotating frame,  much  like a Jacobi ellipsoid \citep{Chandrasekhar1969}. Significantly, for the extended warm-hot (temperature $\lesssim 1$ keV) circum-galactic halo surrounding the galaxy 2MASX~J23453268-0449256 (a spiral) which hosts a Mpc scale GRG \citep{Bagchi2014}, the shape of the  X-ray halo was found to be ellipsoidal rather than spherical \citep{Mirakhor2021,Bagchi2024} with jet axis at a steep angle to the major axis of the halo.

The presence of GRGs within rich groups, clusters, and even super clusters appears to be more prevalent than previously assumed, with many serving as the brightest galaxies within their environment \citep{Sankhyayan2024}, which influences the `min' and `maj' cases. In the case of `edge', the jet is propagating at the periphery of the ambient environment, considering the virial radius of less evolved groups and poor clusters to be varying around $\sim 800$ kpc \citep{Hota2011,Stuardi2020,Oei2023a}.

We assigned the pressure distribution $P(x', y', z)$ of the ambient medium following the density distribution (correlated with the adiabatic gas constant $\gamma$), which then generates an isothermal atmosphere of temperature 1.6 keV \citep{Oei2023a}. A representation of pressure values through contours has also been highlighted in Fig.~\ref{Fig:Setup}, indicating the drastic, slow, and relatively constant changes in pressure gradient along the minor, major, and edge configurations, respectively. To maintain the static equilibrium of such a medium, a gravitational profile has been assigned, derived from the hydrostatic equilibrium equation as,
\begin{equation} \label{eq:4}
    \nabla P = \rho \textbf{g}
\end{equation}
where \textbf{g} being the acceleration due to gravity. The three components of \textbf{g} are represented as follows,
\begin{equation}
\begin{split}
    g_{x'} = -\frac{3\beta}{\gamma} \left( \frac{x'-x_0}{a^2}\right) \left(1+\frac{(x'-x_0)^{2}}{a^{2}}+\frac{(y'-y_0)^{2}}{b^{2}}+\frac{(z-z_0)^{2}}{c^{2}}\right)^{-1}\\
    g_{y'} = -\frac{3\beta}{\gamma} \left( \frac{y'-y_0}{b^2}\right) \left(1+\frac{(x'-x_0)^{2}}{a^{2}}+\frac{(y'-y_0)^{2}}{b^{2}}+\frac{(z-z_0)^{2}}{c^{2}}\right)^{-1}\\
    g_{z} = -\frac{3\beta}{\gamma} \left( \frac{z-z_0}{c^2}\right) \left(1+\frac{(x'-x_0)^{2}}{a^{2}}+\frac{(y'-y_0)^{2}}{b^{2}}+\frac{(z-z_0)^{2}}{c^{2}}\right)^{-1}\\
\end{split}
\end{equation}

\subsection{Configurations of the propagating jets} \label{Subsec:Configurations of the propagating jets}

We have introduced the jet into the computational domain by constructing a cylindrical injection zone with a radius $r_j$ of 1 kpc and a length of approximately 3 kpc. The jets are set to propagate along the negative $x$-axis in all simulations. To model the growth factors responsible for generating GRGs that extend beyond 1 Mpc in total length, and considering the extensive computational demands of each run, our simulations are designed to extend from $+10$ kpc to $-710$ kpc along the $x$-direction. Therefore, we modeled only the one-sided propagation of the jet. We inject it into the domain from $(x,\, y,\, z) \equiv (0,\, 0,\, 0)$ in all cases (see, Fig.~\ref{Fig:Setup}). This setup also required us to maintain the jet radius consistent with the resolution of the domain, resulting in 720 grid points in the $x$-direction (a resolution enhancement effect is discussed in Appendix~\ref{AppendixA: Effects of Resolution Enhancement}). Along the other two axes $(y,\,z)$, the domain was varied around the extent of $[-180,\, 180]$ kpc to optimally capture the jet morphology (see, Table~\ref{Tab:Parametric_space}).

To discuss the thermodynamical properties of the jet, we incorporated an underdense, approximately pressure-matched, magnetic, relativistic jet into the domain. The jet is underdense with respect to the ambient medium, with $\rho_j = 10^{-5}\rho_0$, and is slightly over-pressured with the ambient environment (see, Appendix~\ref{Appendix:Observability of Recollimation Shocks}) at the launching zone ($P_j \sim 5.36\times 10^{-12}$ dyn/cm$^2$), following: \citet{Rossi2017,Musoke2020,Mukherjee2020,Giri2023}. A toroidal magnetic field has been introduced along with the jet, configured as follows,
\begin{equation} \label{eq:5}
B_{y} = B_{j}\, r\, {\rm sin(\theta)},\ \ 
B_{z} = -B_{j}\, r\, {\rm cos(\theta)}
\end{equation}

\noindent where, $(r,\, \theta)$ is the polar coordinate in the $(y,\, z)$ plane, perpendicular to the jet propagation axis, and so, $(B_y,\, B_z)$ represent the components of the toroidal magnetic field in the $(y,\, z)$ plane. $B_j$ is a constant and assigns the strength of the magnetic field, which is determined through the jet magnetization parameter $\sigma$ estimated as the ratio of poynting flux to the matter energy flux as,
\begin{equation} \label{eq:6}
\sigma = \frac{B_j^2}{\Gamma^2 \rho_j h_j};\,\,\,\, \rho_j h_j = \frac{5}{2}P_j + \sqrt{\frac{9}{4}P_j^2 +(\rho_j)^2}
\end{equation}

\noindent where, $\rho_j h_j$ being the relativistic enthalpy, estimated for a Taub–Matthews equation of state \citep{Taub1948,Mignone2005}, and $\Gamma$ being the bulk Lorentz factor of the jet. The value of $\sigma$ is set to $0.01$, a value typically assumed at the early stages of jet propagation \citep{Rossi2017,Mukherjee2020}. To ensure that $\nabla  \cdot \textbf{B} = 0$ is maintained in the computational domain always, we employ the divergence cleaning method in our computations, as designed by \citet{Dedner2002}. We conducted these simulations with second-order spatial accuracy, utilizing linear reconstruction and the HLLC Riemann solver \citep{Mignone2006}.

\begin{table*}
\caption{Parameters that are varied in our simulations.}
\begin{center}
\begin{tabular}{ |c|c|c|c|c|c| } 
 \hline
 Simulation& Jet Power ($Q_{\rm j}$) & Jet-flow & Domain ($\times \ L_0$) & Grid & Center of the ambient\\
 Label& (erg/s) & along environment's & $[x],\,[y],\,[z]$ &  & medium $(x_0,\,y_0,\,z_0)\times(L_0)$\\
 \hline
 GRG\_lp\_min & $2.3 \times 10^{44}$ & minor-axis & $[-7.1,\, 0.1]$ & $720 \times 360 \times 360$ & $(0,\, 0,\, 0)$\\ 
  &  &  & $[-1.8,\, 1.8]$ &  & \\
   &  &  & $[-1.8,\, 1.8]$ &  & \\
 \hline
 GRG\_hp\_min & $7.2 \times 10^{44}$ & minor-axis & $[-7.1,\, 0.1]$ & $720 \times 300 \times 300$ & $(0,\, 0,\, 0)$\\ 
  &  &  & $[-1.5,\, 1.5]$ &  & \\
   &  &  & $[-1.5,\, 1.5]$ &  & \\
   \hline
   GRG\_lp\_maj & $2.3 \times 10^{44}$ & major-axis & $[-7.1,\, 0.1]$ & $720 \times 400 \times 400$ & $(0,\, 0,\, 0)$\\ 
  &  &  & $[-2.0,\, 2.0]$ &  & \\
   &  &  & $[-2.0,\, 2.0]$ &  & \\
   \hline
   GRG\_hp\_maj & $7.2 \times 10^{44}$ & major-axis & $[-7.1,\, 0.1]$ & $720 \times 300 \times 300$ & $(0,\, 0,\, 0)$\\ 
  &  &  & $[-1.5,\, 1.5]$ &  & \\
   &  &  & $[-1.5,\, 1.5]$ &  & \\
   \hline
   GRG\_hp\_edge & $7.2 \times 10^{44}$ & edge & $[-7.1,\, 0.1]$ & $720 \times 360 \times 360$ & $(0,\, -6,\, 0)$\\ 
  &  &  & $[-1.8,\, 1.8]$ &  & \\
   &  &  & $[-1.8,\, 1.8]$ &  & \\
   \hline
\end{tabular}

\label{Tab:Parametric_space}
\end{center}
\small
\textbf{Notes.} The table summarizes the parameters varied in five simulation runs, covering a wide range of jet-ambient medium settings. The simulation labels define the characteristics of the runs, with `lp' and `hp' representing low and high powered jets, corresponding to strong FR I and standard FR II jets respectively, and `min' and `maj' indicating the propagation direction of the jet relative to the ambient environment. The domain and grid sizes are chosen to optimally capture the jet morphology. The last column defines the center of the ambient environment, facilitating the setup of the 5th simulation where the jet flows along the `edge'. Here, $L_0$ is the length unit of 100 kpc.
\end{table*}

The jet power at the time of injection is controlled (in our runs) by the parameter $\Gamma$, the bulk Lorentz factor of the jet. For our simulations, $\Gamma$ is set to values of 3 and 5 to represent two different scenarios. The jet power can then be calculated as follows \citep{Mathews2019,Giri2023},
\begin{equation} \label{eq:7}
   Q_j = \pi r_j^2 {\rm v}_j \bigg[ \Gamma (\Gamma - 1)\rho_j c^2 + \frac{\gamma}{\gamma - 1}\Gamma^2 P_j + \frac{B_j^2}{4\pi}\bigg] 
\end{equation}
Here, ${\rm v}_j$ is the jet propagation speed during its injection into the domain (estimated from $\Gamma$ values). The adiabatic index $\gamma$ varies between $5/3$ and $4/3$ depending on the temperature of the medium, as prescribed by the Taub-Matthews equation of state. 

The evaluated jet power is approximately $2.3 \times 10^{44}$ erg/s for $\Gamma = 3$ and $7.2 \times 10^{44}$ erg/s for $\Gamma = 5$. It is important to note that the radio power (at 1.4 GHz) of an evolved jet system is typically lower than the mechanical power of the jet by a factor of 3 to 300, as highlighted in studies by \citet{Birzan2004} and \citet{Rossi2017}. Based on the models of \citet{Hardcastle2018}, the radio power at 150 MHz can also be estimated from the jet's kinetic power. Relying on these estimates, the two different jet scenarios can then be enumerated as follows \citep[see also,][]{Mingo2019,Dabhade2020a},
\begin{enumerate}[leftmargin=*,labelindent=\parindent]
\item Low jet-power (`lp'):
\begin{itemize}
\item $Q_j = 2.3 \times 10^{44}$ erg/s; \\
Expected radio power resembling powerful FR I sources
\end{itemize}
\item High jet-power (`hp'):
\begin{itemize}
\item $Q_j = 7.2 \times 10^{44}$ erg/s; \\
Expected radio power resembling standard FR II sources
\end{itemize}
\end{enumerate}

\noindent A combination of jet power and ambient medium configuration results in five distinct simulation settings, as highlighted in Table~\ref{Tab:Parametric_space} and schematized in Fig.~\ref{Fig:Setup}. Although GRGs are predominantly of FR II nature, the existence of $\sim 5-8$ percent GRGs with FR I characteristics \citep{Dabhade2020a,Simonte2024}, necessitates verification of the applicability of hypothesized models in this regime of low-powered jet propagation, which is why the choice of `lp' above. These simulations therefore possess the potential to evaluate the effectiveness of growth factors, such as jet power or ambient medium hindrance, in forming the giant radio galaxies through comparative analysis. To note, due to the limited availability of tailed GRGs \citep{Andernach2021}, and the tendency of FR I sources to thrive in rich cluster or dynamically evolved group mediums \citep{Zirbel1997,Mingo2019}, we have not explicitly considered simulating an FR I jet propagating through the outskirts of the environment. Additionally, the interstellar medium (ISM) of the host galaxy has not been specifically modeled here, as our study focuses on sources with jetted structures extending beyond one megaparsec, representing a late-stage evolution. The impact of the ISM on the early phases of such jetted structures will be addressed in a separate study.

All of our simulations have been performed in dimensionless units, with conversions to physical scales defined by the following units: length ($L_0 = 100$ kpc), velocity ($c = 2.99\times 10^{10}$ cm/s), and density ($\rho_0 = 0.001$ amu/cc). Units for other variables can be derived from these three fundamental units. For instance, the unit of time can be calculated as the ratio of the unit length to the unit velocity, resulting in $T_0 = 0.326$ Myr. Finally, we note that two standard passive scalar tracers were injected into the domain: \( \text{tr1} \) to identify jet material and \( \text{tr2} \) to identify ambient material, with values ranging between 0 and 1.

\section{GRG properties and Observational relevance} \label{Sec:GRG properties and Observational relevance}

This section provides a detailed overview of the findings of this work, addressing several key aspects: the morphological characteristics of the evolved structures and their relevance to observed sources, the implications for the ages of such giant jetted structures, and an examination of the (magneto-hydro)dynamical properties to determine whether GRGs are similar to or distinct from their smaller counterparts.

\subsection{Morphological Distinction} \label{Subsec:Morphological Distinction}

\subsubsection{Low-powered jet along the minor axis of the environment} \label{Subsubsec:Low-powered jet along the minor axis of the environment}

\begin{figure*}
\centering
\includegraphics[width=1.8\columnwidth]{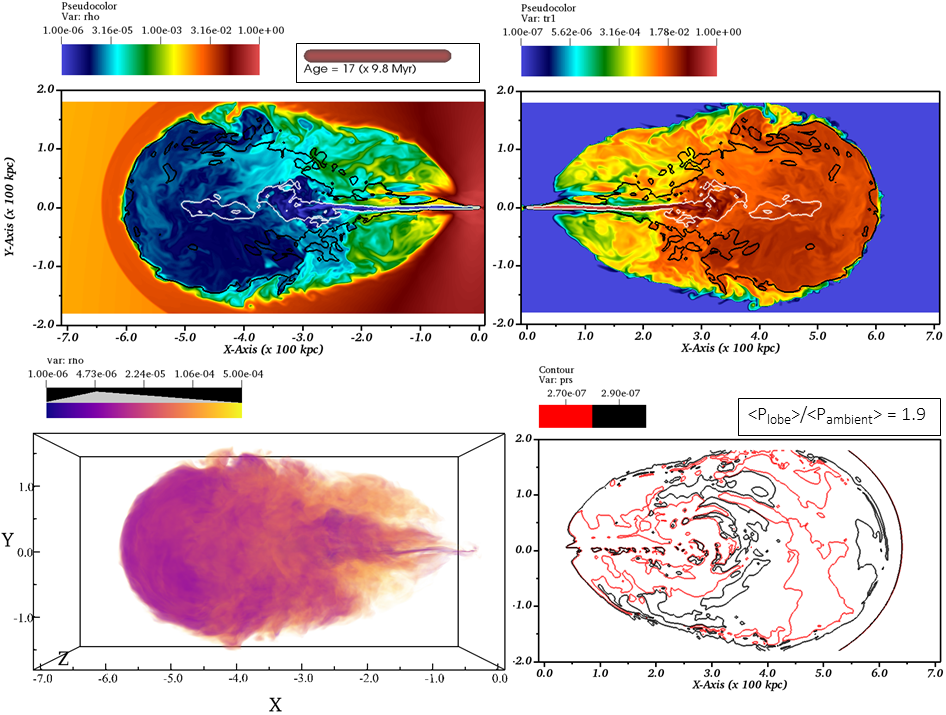}
\caption{Simulation `GRG\_lp\_min' (low-powered jet propagating along the minor axis of the environment), showcasing the structure of the evolved giant radio galaxy at the highlighted age. \textit{Top-left} panel: $x-y$ variation of $(\rho/\rho_0)$, accompanied by contours of velocity ($0.1c$ for white, $0.01c$ for black). \textit{Top-right} panel: tracer distribution showing regions with jet material and their associated fraction, where the highest colorbar value indicates a grid cell completely filled with jet material (flipped for better visualization), with contours representing the same velocities as earlier. \textit{Bottom-left} panel: 3D representation of $(\rho/\rho_0)$ of the lobe, providing finer details with an extra dimension. \textit{Bottom-right} panel: showcasing contours of pressure values $(P_{\rm lobe}/P_0)$, indicating a minimal variation inside the cocoon (flipped for better visualization). The cocoon also appears to be over-pressured compared to the ambient medium, as indicated by the ratio above the subplot. Here, $\rho_0 = 0.001$ amu/cc, $P_0 = 1.5\times10^{-6}$ dyn/cm$^2$, and length represented compared to $L_0 = 100$ kpc. Detail in Section~\ref{Subsubsec:Low-powered jet along the minor axis of the environment}.}
\label{Fig:GRG_lp_min} 
\end{figure*}

The cocoon topology resulting from the propagation of a low-powered jet along the minor axis of the ambient medium at the late stage of its evolution is presented in Fig.~\ref{Fig:GRG_lp_min} (`GRG\_lp\_min'). After an extended evolution period of approximately 166 Myr, a fat lobe has formed, measuring nearly 600 kpc in length and 360 kpc in width. The broad lobe is best recognized in the tracer plot \citep[tracer values, tr1 $\geq 10^{-7}$;][]{Mukherjee2020}, shown in the \textit{top-right} panel of Fig.~\ref{Fig:GRG_lp_min}. This panel also clearly illustrates the velocity distribution of cocoon material within the lobe via contours, which vary between $[0.01c,\, 0.1c]$, indicating that the lobe constituents are evolving sub-relativistically. The jet ridge-line appears to be mildly relativistic even after propagating over such large distances. Since we have simulated only one-sided jet propagation, the total extent of the structure is expected to reach 1.2 Mpc, thus easily qualifying as a giant radio source. 

The morphology of the cocoon is well captured in the 3D volume-rendered plot (\textit{bottom-left} of Fig.~\ref{Fig:GRG_lp_min}), with density thresholds and color transparency applied as represented by the color bar.
This diagram illustrating how the propagating jet transitions into a decollimated, broad flow (around 250 kpc), resulting in the formation of such an extended lobe. Similar transition zones in low-powered jets have also been recognized in: \citet{Monceau-Baroux2015,Nawaz2016,Giri2022b}, showcasing numerically the process of generating plumes by a decollimated jet.

The pressure plot in the \textit{bottom-right} of Fig.~\ref{Fig:GRG_lp_min} shows contours of pressure values ($P/P_0$), indicating minimal variation of thermal pressure inside the cocoon. This is due to the lobes being overpressured relative to the ambient medium by a factor of 1.9, which effectively confines the lobe boundary against the ambient material \citep[similar to e.g.,][]{Mathews2019,Giri2023}. This overpressured phase results in the generation of a bow shock, for example, at the forward-end of the lobe, which is clearly visible in the density slice plot, where the ambient material appears compressed \citep[relevant to a low jet-power case of][]{Simionescu2009}.

From Fig.~\ref{Fig:GRG_lp_min}, it is evident that a significant portion of the jet plasma accumulates in the frontal region of the lobe, likely observable in emission maps, as depicted by the 3D density map. The relevance of such structures in giant radio sources can be found in statistical studies such as: \citet{Subrahmanyan1996,Mack1997,Malarecki2013,Andernach2021,Oei2023}. Individual studies on giant lobes have also been conducted, with \citet{Oei2022} suggesting a lower resistance of the ambient medium to the jet, \citet{Cantwell2020} and \citet{Dabhade2022} indicating morphological transitions between jet and the fat-lobe, beside reporting a significantly higher time of evolution for such structures, \citet{Saripalli2009} and \citet{Hota2011} suggesting a preference for the GRGs to propagate along the minor axis of the host medium, and \citet{Oei2023a} pointing to the lowest lobe-to-ambient medium pressure ratio for such extended configurations, which are also the properties relevant to the present simulation (`GRG\_lp\_min').

\subsubsection{High-powered jet along the minor axis of the environment} \label{Subsubsec:High-powered jet along the minor axis of the environment}

\begin{figure*}
\centering
\includegraphics[width=1.8\columnwidth]{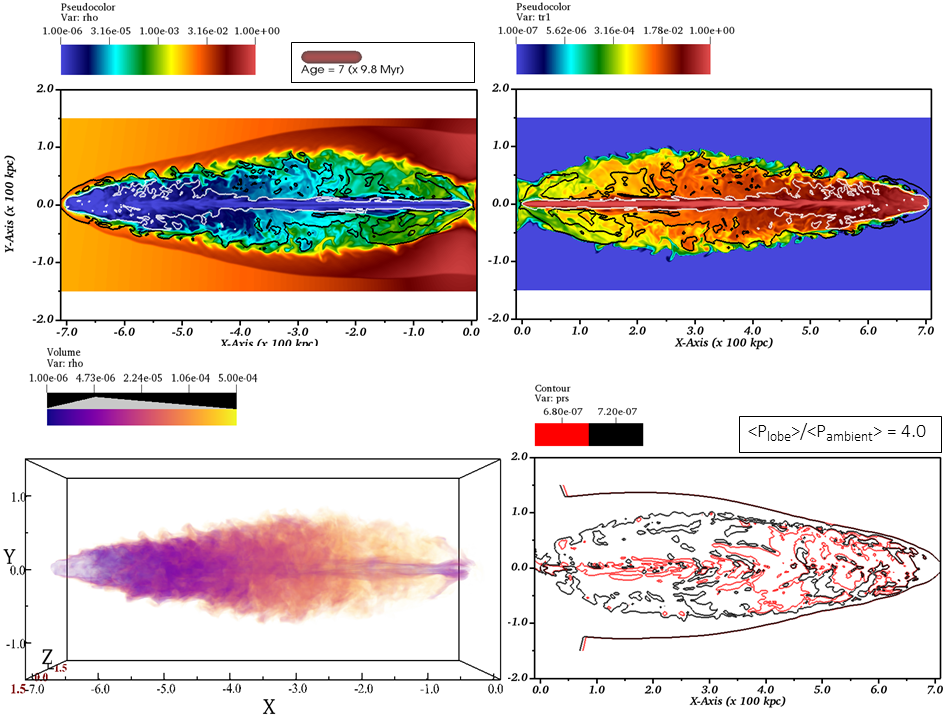}
\caption{A morphological collage similar to Fig.~\ref{Fig:GRG_lp_min}, but for simulation `GRG\_hp\_min' (high-powered jet propagating along the minor axis of the environment), showcasing the structure of the evolved giant radio galaxy at the highlighted age. We emphasize the much shorter evolution time for the jet to reach $\sim 700$ kpc compared to the low-powered jet along the minor axis. Here, $\rho_0 = 0.001$ amu/cc, $P_0 = 1.5\times10^{-6}$ dyn/cm$^2$, and length represented compared to $L_0 = 100$ kpc. Detail in Section~\ref{Subsubsec:High-powered jet along the minor axis of the environment}.}
\label{Fig:GRG_hp_min} 
\end{figure*}

In this scenario (`GRG\_hp\_min'), a high-powered jet propagating along the minor axis of the ambient medium forms a structure shown in Fig.~\ref{Fig:GRG_hp_min} at nearly 68 Myr. The jet-cocoon structure appears thin, with an axial ratio (length to width) of 5.5, similar to the GRG cases highlighted in \citet{Subrahmanyan1996}. We emphasize the significantly shorter evolution time for the jet to reach 700 kpc compared to a low-powered jet along the same axis. With a jet power increased by a factor of 3, compared to the earlier case, the time required for the lobe to extend to 700 kpc is reduced by $\sim 2.5$ times, indicating a substantial impact on the jet evolution time. This suggests that lobe age, in conjunction with radio power and morphology, can serve as a proxy for understanding GRG formation models. 

The frontal part of the lobe still moves rapidly, with speeds $\geq 0.1c$, as seen from the contour plots in the \textit{top-rows} of Fig.~\ref{Fig:GRG_hp_min}. This high speed causes the bow shock to remain attached to the jet head. Despite the high power of the jet, which would typically result in more back-flowing material from the jet head \citep{Cielo2017}, the rapid expansion of the jet leads to a thinner cocoon morphology. However, the combination of substantial back-flow and the lower axial ratio of the cocoon geometry results in higher cocoon pressure, increasing the lobe-to-ambient pressure difference by a factor of 4.0. This is illustrated in the pressure plot in the \textit{bottom-right} of Fig.~\ref{Fig:GRG_hp_min}, which also shows the formation of back-flowing plasma from the jet ridgelines, likely around the recollimation shocks.

The \textit{bottom-left} subplot of Fig.~\ref{Fig:GRG_hp_min} showcases a 3D volume-rendered configuration, highlighting the formation of prominent recollimation shocks (see also the discussion in Appendix~\ref{Appendix:Observability of Recollimation Shocks}). The cocoon develops Kelvin-Helmholtz instabilities at the jet-ambient medium interface, a typical characteristic of kinetically dominated jets \citep{Acharya2021,Giri2022b}. However, these instabilities only marginally hinder the growth of the structure, and the jet head yet remains well-collimated, enabling rapid propagation through the denser environment. It is noteworthy that this jet flows from the center of the ambient medium, suggesting that GRG formation from Brightest Cluster Galaxies is theoretically plausible. This corroborates the findings of a recent study by \citet{Oei2024}, which suggests that luminous GRGs are more likely to be found in denser environments. However, the lower occurrence of such cases (BCG-GRGs) may be due to the general tendency of FR II jets to avoid residing in evolved clusters or rich groups \citep{Gendre2013,Joshi2019}. With the one-sided lobe length of the simulated structure exceeding 700 kpc, the total extent is expected to reach 1.4 Mpc, thereby comfortably meeting the definition of a GRG in the literature.

The availability of GRG morphologies such as this is commonly observed in statistical studies. A powerful jet with thin cocoon geometry and hotspots at the lobe ends can be observed in studies such as: \citet{Stuardi2020,Dabhade2020a,Simonte2022,Oei2023}. There have been individual studies on similar morphological sources, such as \citet{Subrahmanyan2008}, indicating FR II GRGs expecting to have well-bounded lobes if not a relic; \citet{Jamrozy2008}, highlighting a lower spectral age for such sources; \citet{Delhaize2021}, showing the recollimation shock knots, and \citet{Machalski2011b}, suggesting ballistic propagation of the jet with a younger age of evolution.

\subsubsection{Low-powered jet along the major axis of the environment} \label{Subsubsec:Low-powered jet along the major axis of the environment}

\begin{figure*}
\centering
\includegraphics[width=1.8\columnwidth]{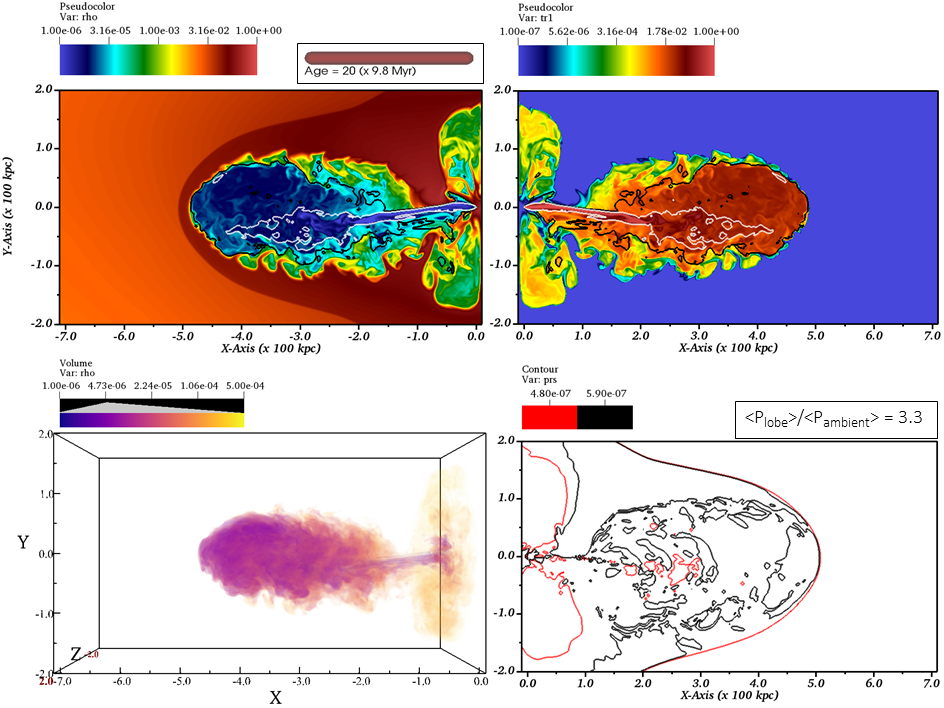}
\caption{A collage similar to Fig.~\ref{Fig:GRG_lp_min}, but for simulation `GRG\_lp\_maj' (low-powered jet propagating along the major axis of the environment), showcasing the structure of the evolved giant radio galaxy at the highlighted age. The structure has evolved to only 500 kpc on one side ($\sim 1$ Mpc in total extent) over a longer period of evolution, indicating higher jet frustration imposed by the environment while traveling along the major axis. The emerged structure is known as an X-shaped radio galaxy, and given its giant size, it is typically recognized as a GRG-XRG. Here, $\rho_0 = 0.001$ amu/cc, $P_0 = 1.5\times10^{-6}$ dyn/cm$^2$, and length represented compared to $L_0 = 100$ kpc. Deatil in Section~\ref{Subsubsec:Low-powered jet along the major axis of the environment}.}
\label{Fig:GRG_lp_maj} 
\end{figure*}

When a low-power jet propagates along the major axis of the ambient environment (`GRG\_lp\_maj'), it encounters higher jet disruption compared to other scenarios in our simulations, resulting in a shorter jet length over a longer evolution time. The one-sided lobe length reaches approximately 500 kpc (see, Fig.~\ref{Fig:GRG_lp_maj}), indicating a total extent of 1 Mpc, qualifying it again as a giant radio galaxy. Propagating along a path of lowest pressure gradient (Fig.~\ref{Fig:Setup}), the higher opposing force to the jet propagation leads to a complex morphology, known as X-shaped radio galaxies \citep[e.g.,][]{Mahatma2023a}. The formation of such sources is primarily believed to result from the back-flow of plasma from the jet head, which gets diverted along the minor axis of the ambient medium, a path with the highest pressure gradient that aids the backflowing plasma to follow that path, forming the structure known as a wing \citep{Capetti2002,Hodges-Kluck2011,Rossi2017,Giri2022a}. Therefore, the wing typically consists of the older jet plasma and lies at a large angle to the active lobe \citep[for a review, see][]{Giri2024}. Giant X-shaped radio galaxies are recognised as GRG-XRGs, examples of which can be found in the recent past as well \citep{Saripalli2009,Cotton2020}, which have been found to support the classic back-flow scenario. 

The active lobe evolves mostly with sub-relativistic speed, as expected (Fig.~\ref{Fig:GRG_lp_maj}). As the jet grows in size, it appears to become disconnected from the back-flowing material supply to the wing, which is evident in the 3D volume rendered plot. This phenomenon may indicate the observation highlighted by \citet{Saripalli2009} that the larger the extent of the sources along the active lobe, the smaller the wing to lobe length ratio becomes. The lobe edges are susceptible to Kelvin-Helmholtz instabilities, leading to the mixing of ambient material with cocoon material. The jet flows along the major axis of the ambient medium with similar internal energy input as the case `GRG\_lp\_min', which is now stored within a smaller active lobe volume, causing higher internal pressure. Consequently, the lobe becomes overpressured relative to the ambient environment by a factor of 3.3. It is interesting to mention that the wing structure has marginally lesser internal pressure than the active lobe, but still remains overpressured compared to the ambient medium, promising its slower but further growth \citep[see the back-flowing case of][]{Giri2023}. Given the substantial backflow reaching the center, the hot plasma is anticipated to thermally influence both the interstellar medium and the nuclear region \citep[e.g.,][]{Landt2010}.

It's worth noting that the time required for the jet to form a lobe extending 500 kpc on one side is 196 Myr for the highlighted simulation setting. This represents a considerably longer dynamical evolution time, several times higher than the spectral ages typically associated with GRGs \citep[see, for a review,][]{Dabhade2023}. While the dynamical age can vary by a factor of 1 to 5 compared to the spectral age for GRGs \citep{Machalski2009}, the obtained dynamical age of the formed structure here is substantially long. Therefore, as previously mentioned, combining the age of evolution of the GRG with its radio power and morphology may allow for the disentanglement of formation models.

The occurrence of X-shaped radio galaxies themselves comprises only $5-10$ percent of known radio galaxies \citep{Leahy1992,Yang2019}, suggesting that GRG-XRGs may be even less common. However, this fraction may be influenced by the sensitivity limits of radio telescopes, as demonstrated by \citet{Yang2019}. With significant improvements in current and upcoming radio telescopes, the discovery rate of such systems is expected to increase dramatically \citep[see, e.g.,][]{Mahatma2023}. At present, examples of GRG-XRGs can be found in statistical studies conducted by: \citet{Saripalli2009,Dabhade2020b,Bruni2021}. While the later studies focuses on modeling the morphological appearances of these systems, the former showcases how the propagation axis of the jet in GRGs with wings tends to align along the major axis of the host galaxy. Given that the large-scale environment of the majority of XRGs follows the geometry and configuration of the host galaxy \citep{Hodges-KLuck2010}, it is expected that GRGs with wings would also propagate along the major axis of their large-scale environment, similar to what our simulation result indicates. In general, off-axis lobe distortions forming wings in giant radio galaxies are not rare, as observed in studies by \citet{Subrahmanyan1996} and \citet{Malarecki2013}.

It is important to note that the formation of X-shaped radio galaxies may not solely result from classical back-flow scenarios. Other mechanisms, such as binary SMBH mergers, unstable mass accretion by the SMBH, or galaxy mergers causing rotation in the ambient environment that shifts the jet propagation direction, have also been reported to produce winged galaxies \citep[see, review by:][]{Gopal-Krishna2012,Giri2024}. This suggests that, while not all (GRG-)XRGs may be attributed to back-flow processes, this mechanism however could explain majority of cases due to its naturally occurring nature \citep[similar to the conclusions of, e.g.:][]{Saripalli2009,Joshi2019,Bruni2021,Dabhade2022a}.

\subsubsection{High-powered jet along the major axis of the environment} \label{Subsubsec:High-powered jet along the major axis of the environment}

\begin{figure*}
\centering
\includegraphics[width=1.8\columnwidth]{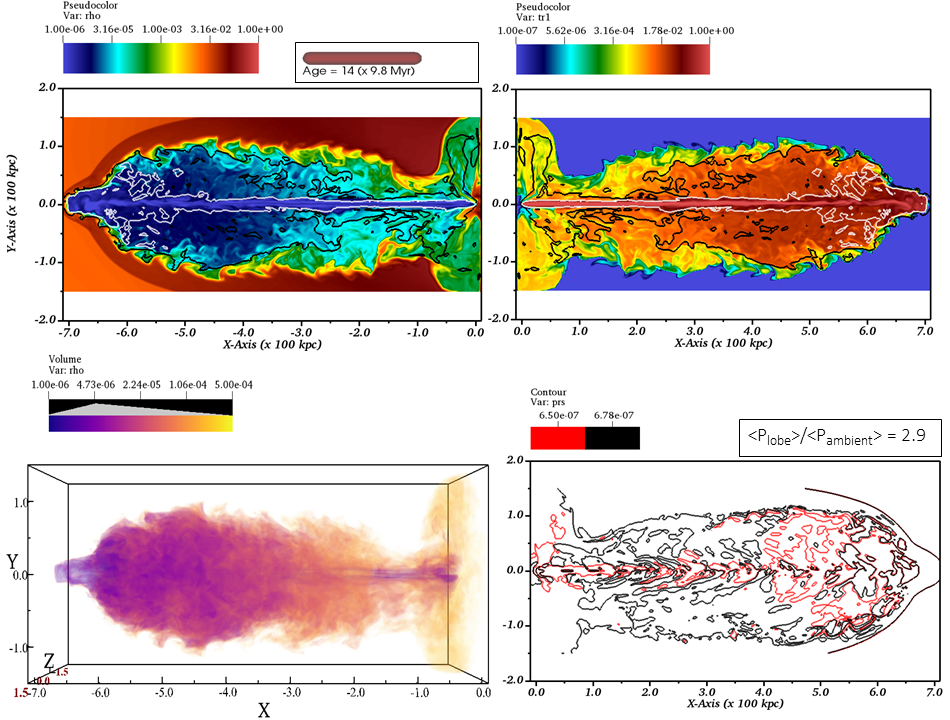}
\caption{A collage similar to Fig.~\ref{Fig:GRG_lp_min}, but for simulation `GRG\_hp\_maj' (high-powered jet propagating along the major axis of the environment), showcasing the structure of the evolved giant radio galaxy at the highlighted age. This morphology exhibits the formation of a mini-wing perpendicular to the active lobe. Additionally, it demonstrates the phenomenon of rapid jet propagation (top-hat jet front), leaving the cocoon behind at the given time. Such topology are not uncommon for giant radio galaxies and have been observed in many instances. Here, $\rho_0 = 0.001$ amu/cc, $P_0 = 1.5\times10^{-6}$ dyn/cm$^2$, and length represented compared to $L_0 = 100$ kpc. Detail in Section~\ref{Subsubsec:High-powered jet along the major axis of the environment}.}
\label{Fig:GRG_hp_maj} 
\end{figure*}

A high-powered jet can overcome the increased jet frustration imposed by the environment when propagating along its major axis, as seen in the `GRG\_hp\_maj' case, resulting in the evolved structure shown in Fig.~\ref{Fig:GRG_hp_maj} at 137 Myr. This structure resembles the earlier scenario of `GRG\_lp\_maj' (Fig.~\ref{Fig:GRG_lp_maj}), but with a shorter secondary lobe perpendicular to the active lobe, identified as mini-wings \citep[primary to secondary lobe length ratio $< 0.8$;][]{Cheung2007}. The reason for the shorter wing is that, with the increase in jet speed, the growth rate of the cocoon length also accelerates. Consequently, a substantial amount of the back-flow material struggles to reach the center and the wing, despite the faster production of back-flowing plasma for a high-speed jet. This phenomenon is numerically demonstrated by \citet{Rossi2017} and supported by observations \citep[e.g.,][]{Gillone2016}, indicating that only low-powered jets ($\sim$ FR I/II boundary) propagating along the lowest pressure gradient path of the ambient medium (around the major axis) can generate wings from the over-pressured back-flow model. The observational existence of such mini-winged GRGs has also been reported by \citet{Saripalli2009}, who inferred that the more extended the source, the shorter the wings it possesses. A handful of similar sources have been identified in statistical studies by \citet{Malarecki2013} and \citet{Stuardi2020}. Considering that the one-sided active lobe length reaches 700 kpc, the total extent of this source is expected to be 1.4 Mpc, thereby qualifying as a prominent giant radio galaxy with lobe-to-ambient medium thermal pressure ratio of 2.9 (\textit{bottom-right} of Fig.~\ref{Fig:GRG_hp_maj}).

The morphology also reveals a top-hat-like feature at the lobe front, best captured in the volume rendered image, indicating that the jet has broken out of the cocoon due to its propagation speed (Fig.~\ref{Fig:GRG_hp_maj}). While this prominent feature is not consistently present throughout the course of its evolution, the existence of an arrowhead-like structure in the lobe front is evident, representing the rapid propagation of this high-powered jet, despite it being a major-axis source. 

Such extended jet head features have been observed in GRGs as well, exemplified by: \citet{Saripalli1994,Safouris2009,Bruni2021}, who also have documented the presence of multiple recollimation shocks, similar to our's (see, volume-rendered plot; \textit{bottom-left} of Fig.~\ref{Fig:GRG_hp_maj}; and also Appendix~\ref{Appendix:Observability of Recollimation Shocks}). \citet{Chen2018} have also shown the evidence of such jet-head extension along with the formation of fat lobe in the frontal part of the cocoon, due to the encounter of higher jet obstruction along its propagation direction as inferred from the galaxy distribution (similar to what we report). Another example could be found in the work of \citet{Oei2022}, which illustrates a fat frontal cocoon attached to the top-hat jet structure. However, the estimated jet power derived from radio observations appears to be on the lower side, possibly due to non-negligible radiative cooling given the source's size and redshift. These subtle features, combined with the lobe configuration, are challenging to identify in statistical studies unless the GRG morphology is highly resolved; however such structures are not rare \citep{Simonte2024,Dabhade2020a}.

\subsubsection{High-powered jet along the edges of the environment} \label{Subsubsec:High-powered jet along the edges of the environment}

\begin{figure*}
\centering
\includegraphics[width=1.8\columnwidth]{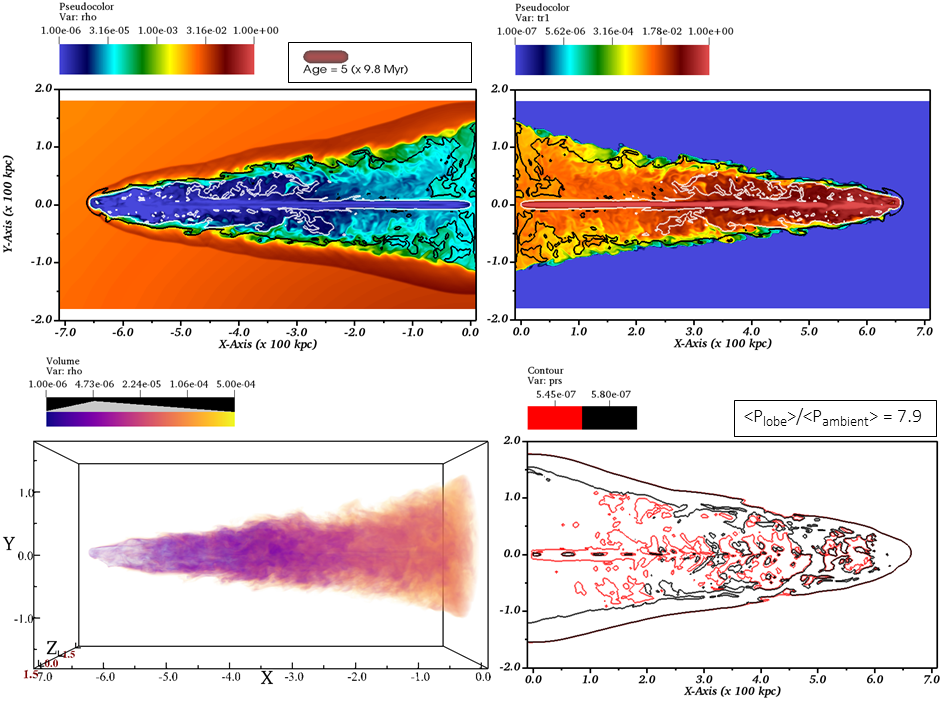}
\caption{A collage similar to Fig.~\ref{Fig:GRG_lp_min}, but for simulation `GRG\_hp\_edge' (high-powered jet propagating along the edges of the environment), showcasing the structure of the evolved giant radio galaxy at the highlighted age. The dynamical age of 49 Myr is considerably low for a GRG and comparable to smaller radio galaxies, hinting at the potential formation of even larger-scale GRGs within a satisfactory time frame compared to radiative ages. Here, $\rho_0 = 0.001$ amu/cc, $P_0 = 1.5\times10^{-6}$ dyn/cm$^2$, and length represented compared to $L_0 = 100$ kpc. Detail in Section~\ref{Subsubsec:High-powered jet along the edges of the environment}.}
\label{Fig:GRG_hp_edge} 
\end{figure*}

In the `GRG\_hp\_edge' scenario (Fig.~\ref{Fig:GRG_hp_edge}), the jet propagates in a low-density environment along the edges of the tri-axial ambient medium. This results in the jet encountering less resistance compared to the other models considered here. Combined with the high power of the jet, the encountered drag against jet travel is negligible, allowing for rapid jet growth. Consequently, the jet reaches a distance of approximately 650 kpc (with a total expected extent of 1.3 Mpc) in just 49 Myr, a dynamical age comparable to that of smaller radio galaxies \citep{Harwood2017}. 

Considering the arrowhead-like structure at the jet head, it can be inferred that the jet is propagating at a faster rate even after covering such distances. If the jet propagation continues unhindered for a considerably longer period, it could cover an enormous length-scale, explaining the rarer giant sources with lengths extending beyond 3 Mpc \citep{Andernach2021}. As mentioned earlier for the case `GRG\_hp\_min', the combination of a high-power jet producing substantial back-flowing material, along with a thinner cocoon due to faster jet propagation, generates significantly higher lobe pressure. Compared to the pressure of the ambient medium at the edges, the lobe-to-ambient environment pressure difference becomes a factor of 7.9. Being high-powered, the recollimation shocks are quite evident in the pressure plot (\textit{bottom-right} of Fig.~\ref{Fig:GRG_hp_edge}; see also the discussion in Appendix~\ref{Appendix:Observability of Recollimation Shocks}).

A detailed examination of the lobe morphology reveals interesting features such as the formation of a broad cocoon near the jet-ejection zone, with the cocoon extending more towards the positive $y$-axis, i.e., towards the outskirts of the ambient medium. This asymmetry is caused by two factors. First, the high-speed back-flowing plasma (as shown in the velocity contours in the \textit{top} panel of Fig.~\ref{Fig:GRG_hp_edge}) allows the jet material to reach the center rapidly. Then, as the material starts accumulating, the buoyancy force (acting outward from the ambient medium, in the direction of the pressure gradient) generates this asymmetric extension. Such buoyancy effects have been observed in smaller radio sources \citep[e.g.,][]{Leahy1984}. However, for GRGs, we suspect observing such cocoon structures may be difficult as the extended zone consist of older, cooled jet plasma \citep[see, PKS 2356-61 for a contrary;][]{Ursini2018}. Nonetheless, morphological structures similar to the frontal part of the lobe have been recognized in many statistical studies \citep{Bruggen2021, Dabhade2020a, Stuardi2020, Simonte2022, Oei2023}. Individual studies, such as \citet{Machalski2008} referred the existence of an exceptionally low ambient medium density, \citet{Delhaize2021} identifying a similarly shaped but bent source likely evolving in the outskirts of a poor group, and \citet{Konar2004,Sebastian2018} suggesting a considerably low age and similar jetted morphology, implying our simulated scenario at play.

\subsection{Temporal Signatures} \label{Subsec:Temporal signatures}

\subsubsection{Evolution of axial ratios (length to width)} \label{Subsubsec2:Evolution of axial ratios}

The ratio of the length to the width of the cocoon has been used for giant radio galaxies to investigate whether this parameter (axial ratio) can provide insights into distinguishing GRGs from their smaller counterparts. For example, studies by \citet{Subrahmanyan1996,Saripalli1996} have shown that there is no significant difference in axial ratios between GRGs and SRGs. Furthermore, a detailed understanding of the axial ratio geometry of the radio lobes aids in accurately estimating dynamical quantities \citep{Hardcastle2018}, particularly concerning GRGs \citep{Subrahmanyan2008, Machalski2008,Machalski2009}. Therefore, we have plotted the axial ratio distribution along the length of the cocoon for evolved systems as tracked by the tracer plots in Fig.~\ref{Fig:GRG_lp_min}-\ref{Fig:GRG_hp_edge}. Additionally, we have plotted the temporal evolution of this parameter, providing better insights into the correlation with their earlier phases.

\begin{figure*}
\centering
\includegraphics[width=1.98\columnwidth]{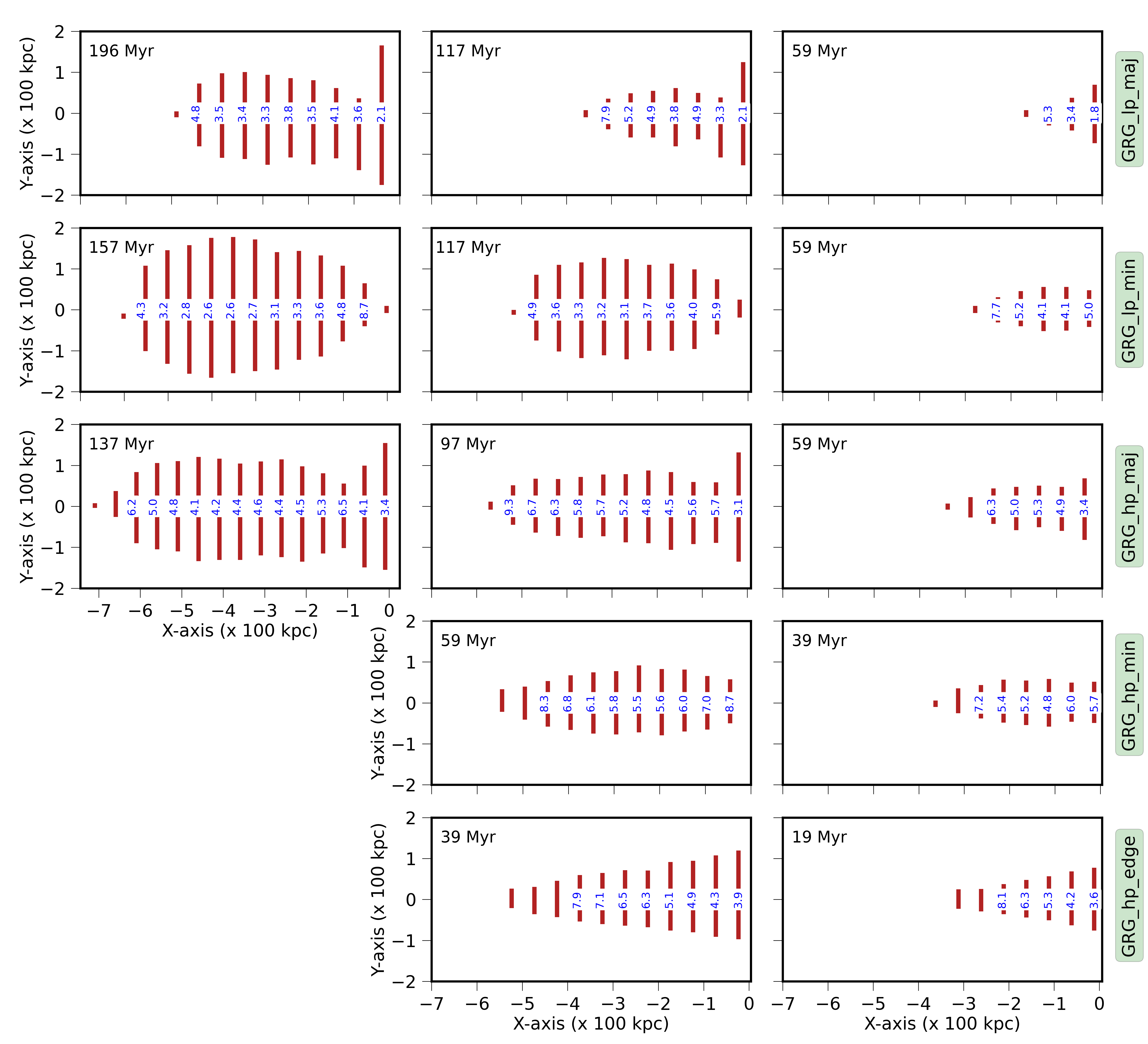}
\caption{Demonstration of temporal evolution of the structural topology originated from five simulations conducted in this work. Each row presents a simulation case, with labels attached to the rightmost part of that row. The evolution of the sources starts from the right column and progresses leftward, with the corresponding ages mentioned in the top corner of each box. The red-lines represent lateral extent of the radio lobe, i.e., the width of the lobe, and the numbers attached to the lines denote the axial ratios (ratio of total length to width at that time) spread across the cocoon.}
\label{Fig:GRG_RT} 
\end{figure*}

This is depicted in Fig.~\ref{Fig:GRG_RT}, where the lobe of the GRG is delineated by tracer values, tr1 $\geq 10^{-7}$ \citep{Mukherjee2020}. For cases where the jet propagates rapidly (`GRG\_hp\_min' and `GRG\_hp\_edge'), the cocoon exhibits similar evolutionary patterns in both their giant and smaller phases, indicating a self-similar expansion (bottom two panels of Fig.~\ref{Fig:GRG_RT}). As mentioned earlier, morphological similarities with GRG structures akin to these two cases have been observed, and a few studies have also reported axial ratio values for such cases that resemble our findings \citep[e.g.,][]{Machalski2008}. In this context, the study by \citet{Andreasyan1999} emphasized that elongated radio sources with FR II-type jet power tend to preferentially orient themselves along the minor axis of the host galaxy, which aligns with the jet-ambient medium configuration applied for these two (low-jet frustration) cases.

Considering the cases `GRG\_lp\_maj' and `GRG\_hp\_maj', both produce similar morphology with wings. The geometry of the active lobe, inferred from the axial ratios, resembles a typical cylinder with minimal variation along its length. Due to higher jet frustration in the low-powered jet, the axial ratio values are slightly smaller, indicating a fatter lobe (with respect to length covered) compared to the high-powered jet. The jet with low-power also produces relatively longer wings than the high-powered case, as reflected by the axial ratio. Compared to their earlier phases with smaller lengths, a mild deviation from self-similar evolution is observed, where increase of width near the frontal part of the lobe has been observed over time. Such axial ratio values are not however uncommon in GRGs, as highlighted by \citet{Subrahmanyan1996}.

The run `GRG\_lp\_min' represents the case with the lowest jet power along the minor axis of the ambient medium, generating a fat lobe with corresponding axial ratio values as highlighted in Fig.~\ref{Fig:GRG_RT}. The morphology also deviating considerably from a cylinder of radius similar to the maximum width of the lobe. Furthermore, compared to its earlier phases, a deviation from self-similar evolution has been observed. Considering their axial ratio values, \citet{Machalski2009} noted resembling numbers for several grown GRGs.

Thus, this parameter `axial ratio', serve as an supplementary tool in deciphering the formation processes of GRGs, complementing the conclusions acquired from age, morphology, and radio power of GRGs. A concise overview of this is well represented in Fig.~\ref{Fig:GRG_RT}.

Finally, to assess the deviation in cocoon volume compared to geometric shapes like a cylinder or sphere, we analyzed the cocoon volume fraction. Initially, we determined the volume occupied by the cocoon material for all simulations at the specified final times outlined for each case in Figure~\ref{Fig:GRG_RT}. Then, we calculated the length of the cocoon and the maximum width of the (primary) lobe to compare the actual cocoon volume (${\rm V_c}$) relative to a cylinder ($\rm V_{cyl}$) with the length as cocoon length and maximum width as its diameter, and to a sphere (${\rm V_{sphr}}$) with a diameter equivalent to the length of the cocoon. The estimations are reported in Table~\ref{Tab:Volume_fraction}. Note that these values are estimated for the one-sided cocoon-structure simulated in our cases.

\begin{table}
\caption{Comparison of different volume measurements.}
\begin{center}
\begin{tabular}{ |l|c|c|c|c| } 
 \hline
 Simulation& Age & ${\rm V_c}$ ($\times 10^6$) & ${\rm V_c}/{\rm V_{cyl}}$ & ${\rm V_c}/{\rm V_{sphr}}$  \\
 (GRG\_) & (Myr) & kpc$^3$ & $(\times 10^{-3})$ & $(\times 10^{-3})$  \\
 \hline
 lp\_maj & 196 & 0.15 & 3.3 & 2.2 \\
 \hline
 lp\_min & 157 & 0.35 & 6.4 & 2.7 \\
 \hline
 hp\_maj & 137 & 0.25 & 7.3 &  1.4\\
 \hline
 hp\_min & 59 & 0.09 & 6.9 & 0.9 \\
 \hline
 hp\_edge & 39 & 0.14 & 6.0 & 1.4 \\
 \hline
\end{tabular}

\label{Tab:Volume_fraction}
\end{center}
\small
\textbf{Notes.} Estimation of (one-sided) cocoon volumes ($\rm V_c$ representing simulation value) for different simulation cases (as noted in column 1) at evolved times similar to those in Fig.~\ref{Fig:GRG_RT} (as also noted in column 2), and their comparison to the equivalent cylinders, as highlighted in column 4 ($\rm V_{cyl}$: length as cocoon length, diameter as cocoon width and spheres, as highlighted in column 5 ($\rm V_{sphr}$: diameter as cocoon length).
\end{table}

\subsubsection{Evolution of lobe lengths (comparison to theory)} \label{Subsubsec:Evolution of lobe lengths}

\begin{figure}
\centering
\includegraphics[width=\columnwidth]{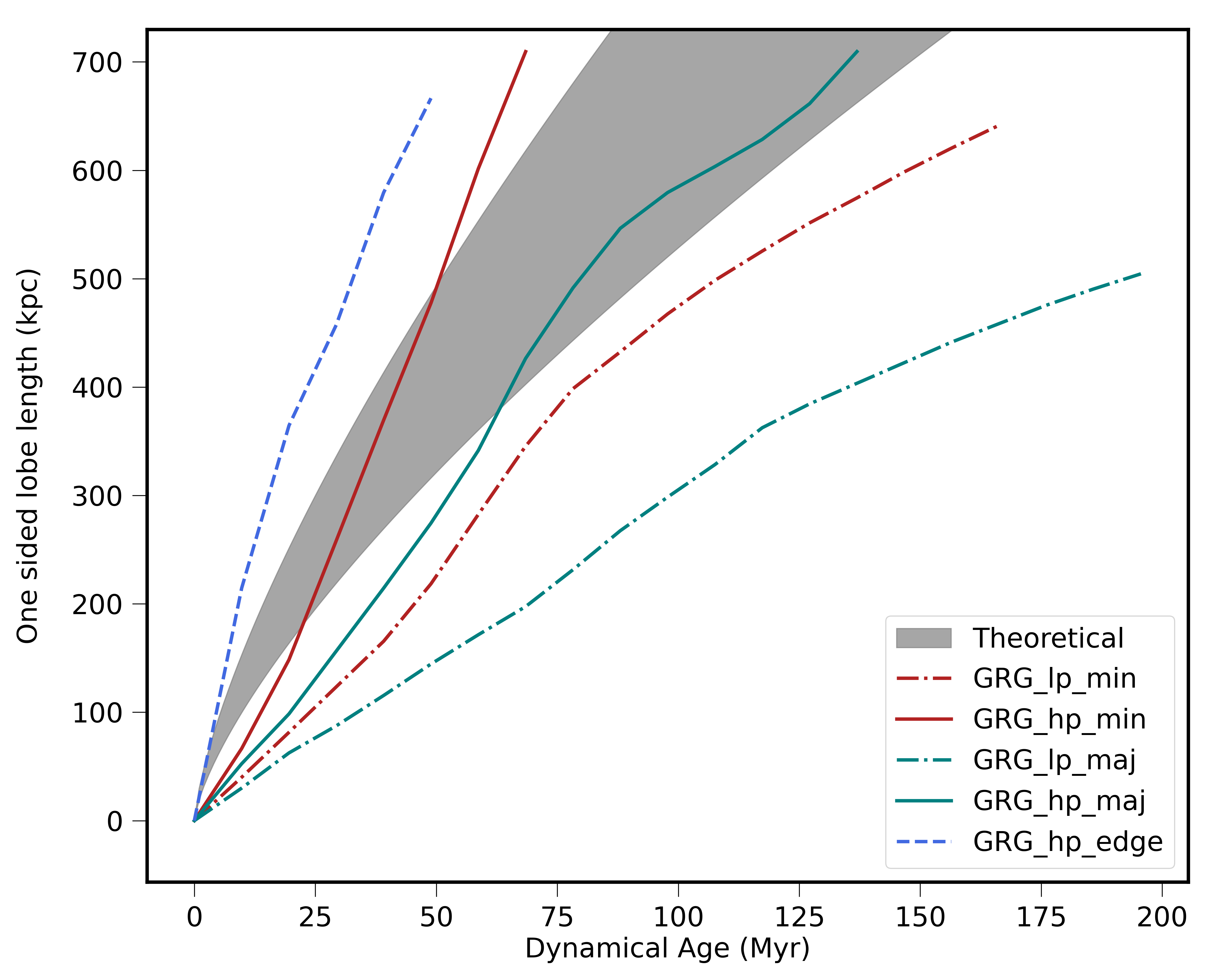}
\caption{Development of cocoon length over the course of evolution for the five simulation cases considered in this work. The shaded region represents the theoretically allowed zones in the length-age diagram for the extended parameter space considered across the simulations. A diverse set of behavior compared to theoretical predictions has been observed here, as discussed in Section~\ref{Subsubsec:Evolution of lobe lengths}. Notably, the blue dashed line represents the case where the jet grows at the outskirts of the ambient medium, thereby, propagating faster than the theoretical predictions.}
\label{Fig:GRG_Length} 
\end{figure}

The continuous variation of the length growth of cocoon over the course of its evolution time has been tracked and presented in Fig.~\ref{Fig:GRG_Length}. As, we are tracking the cocoon geometry, we have used tracer values as, tr1 $\geq 10^{-7}$, similar to \citet{Mukherjee2020}, and then find out the largest cocoon length along the $-x$ direction. Fig.~\ref{Fig:GRG_Length} showcases a diverse set of behaviors consistent with the earlier discussion, with a comparison to theoretical estimates.

To deduce the theoretical framework of cocoon length versus age, we have adopted the formulation of \citet{Kaiser1997}, which has been utilized in deducing the ages of GRGs corresponding to their length by, e.g., \citet{Machalski2009} and \citet{Machalski2011}. They used the geometry of the bow-shock formed from mild to high-powered jets while propagating in a $\beta$-profiled environment, similar to our adopted models. Their analytical calculations highlighted the relationship between the length of the jet, consequently cocoon ($L_j$) and the time of evolution ($t$) as,

\begin{equation} \label{eq:8}
    L_j = c_1\, a_0\, \Bigg( \frac{t}{\tau} \Bigg)^{3/(5 - \beta)};\,\,\,\, \tau = \Bigg( \frac{a_0^5 \, \rho_0}{Q_0} \Bigg)^{1/3}
\end{equation}

\noindent Here, $c_1$ is a dimensionless constant, $a_0$ represents the core radius of the ambient profile which varies with the values of $(a,\, b,\, c)$ in our cases, $\rho_0$ is the density of the ambient medium at the core, and $Q_0$ primarily indicates the kinetic power of the jet. The formulation of $c_1$ is complicated and is dependant on the jet opening angle $\vartheta$ \citep[assumed to be $5^{\circ}$;][]{Horton2020}, adiabatic indexes for ambient medium, sub-relativistic cocoon and relativistic jet material as $\gamma_x = 5/3$, $\gamma_c = 5/3$ and $\gamma_j = 4/3$ respectively \citep[following Taub–Matthews equation of state;][]{Mignone2005}, and two other constants $c_2,\, c_3$ as,

\begin{align} \label{eq:9}
    c_1 &= \left[ \frac{c_2}{c_3\, \vartheta^2} \frac{(\gamma_x + 1)(\gamma_c - 1)(5-\beta)^3}{18 \left\{ 9 \left[ \gamma_c + (\gamma_c - 1) \frac{c_2}{4\vartheta^2}\right] -4 - \beta \right\}} \right]^{1/(5-\beta)}\\
    c_2 &= \left[ \frac{(\gamma_c - 1)(\gamma_j - 1)}{4\, \gamma_c} + 1\right] ^{\gamma_c/ (\gamma_c - 1) } \frac{\gamma_j + 1}{\gamma_j - 1}\\
    c_3 &\approx \frac{\pi \, \vartheta^2}{c_2}
\end{align}

\noindent Given the diverse set of parameter choices in our simulation cases, we evaluated the theoretically allowed range of values and highlighted them with a shaded region in Fig.~\ref{Fig:GRG_Length}. 

Figure~\ref{Fig:GRG_Length} shows that jets with lower kinetic power lag significantly behind theoretical estimates throughout their evolution, taking substantially longer to reach comparably shorter distances. This discrepancy arises because such jets tend to produce fat lobes by loosing the jet's collimation or by bending, whose lateral extent grows faster over time (consequently growth in jet-hindrance), deviating from self-similar evolution (see Section~\ref{Subsubsec2:Evolution of axial ratios}). In contrast, the theoretical evaluation assumes a self-similar expansion of the cocoon structure, where the jet is assumed to remain collimated. The evolution of such cases has been previously recognized by \citet{Rossi2017} and \citet{Giri2022a}.

The cases with high-powered jets are found to reach the theoretically allowed rate of expansion. As indicated earlier, the faster evolving jet of `GRG\_hp\_min,' which develops self-similar growth, mostly follows the theoretical gray zone. However, in the beginning phases of it, the cocoon's geometric evolution in the dense core of the ambient medium is almost spherical rather than elongated causing the lag. While travelling in the outskirts, the jet's ram pressure and the environment's pressure gradient force accelerate the jet propagation. For `GRG\_hp\_maj,' being a major-axis source, the core's influence remains effective for a longer time, leading to the extended lag in the early phase of its propagation. Once out of the core region, the continuous ram pressure of the jet increases its speed. However, at later stages, the lobe expansion is starting to decreases again, as the cocoon is found to deviate from a self-similar expansion. Such evolutionary behavior of high-powered jets has been recognized previously by \citet{Hardcastle2018}.

For the case of 'GRG\_hp\_edge', the evolving length of the jet or cocoon exceeds the theoretical prediction, as expected. This occurs as the jet propagates at the outskirts of the ambient medium with minimal jet-obstruction compared to the theoretical prediction, which assumes the jet always propagates from the center of the ambient medium. We again note the significantly shorter evolution time required for this jet to cover giant scales, indicating why GRGs prefer to avoid dense environments and why large-scaled GRGs ($\geq 3$ Mpc) evolve in underdense environments \citep{Sankhyayan2024}.

\subsection{(Magneto-hydro)Dynamical Characteristics} \label{Subsec:Dynamical characteristics}

\subsubsection{Lobe pressure and expansion speed} \label{Subsubsec:Lobe pressure and expansion speed}
\begin{figure*}
\centering
\includegraphics[width=\columnwidth]{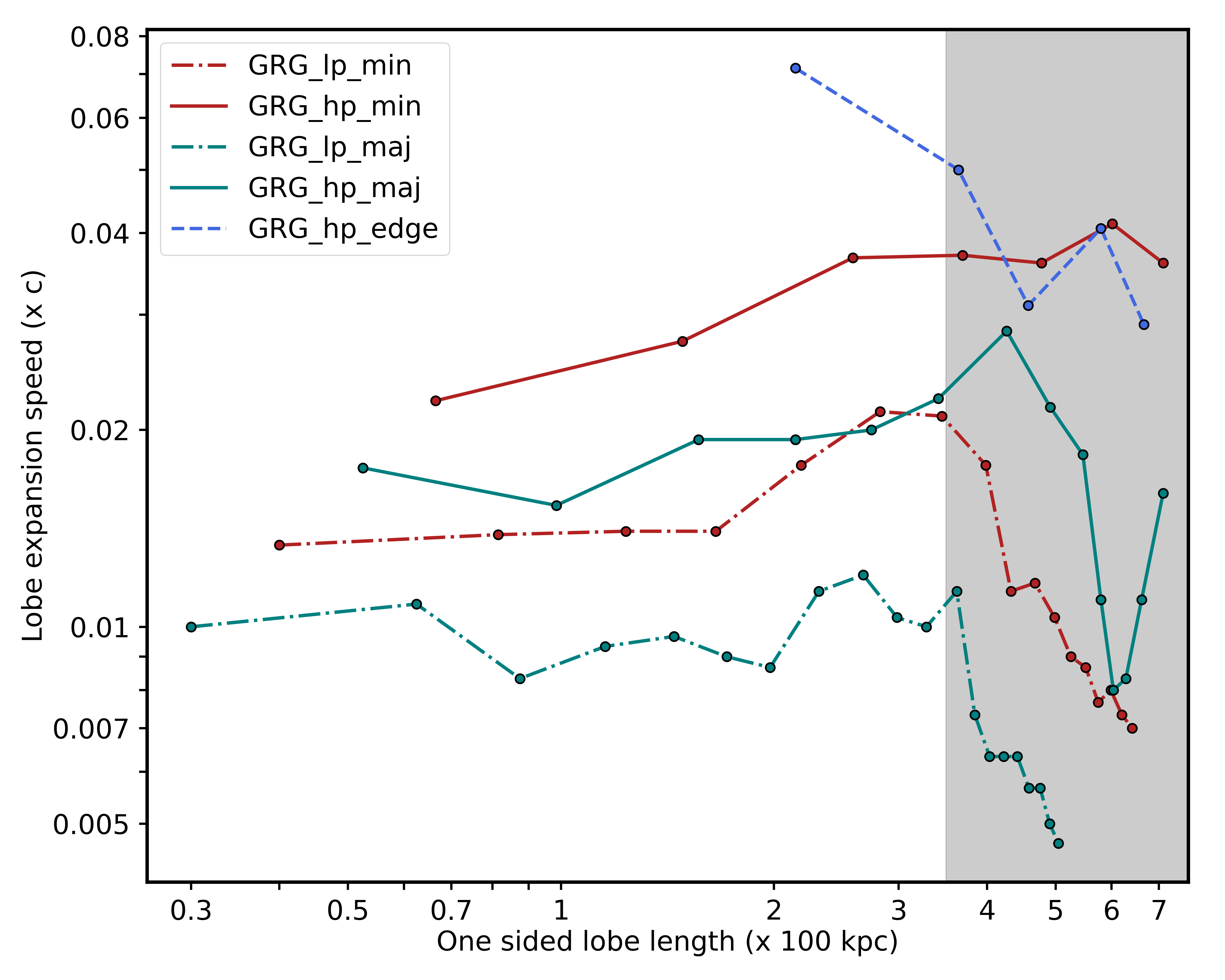}
\includegraphics[width=\columnwidth]{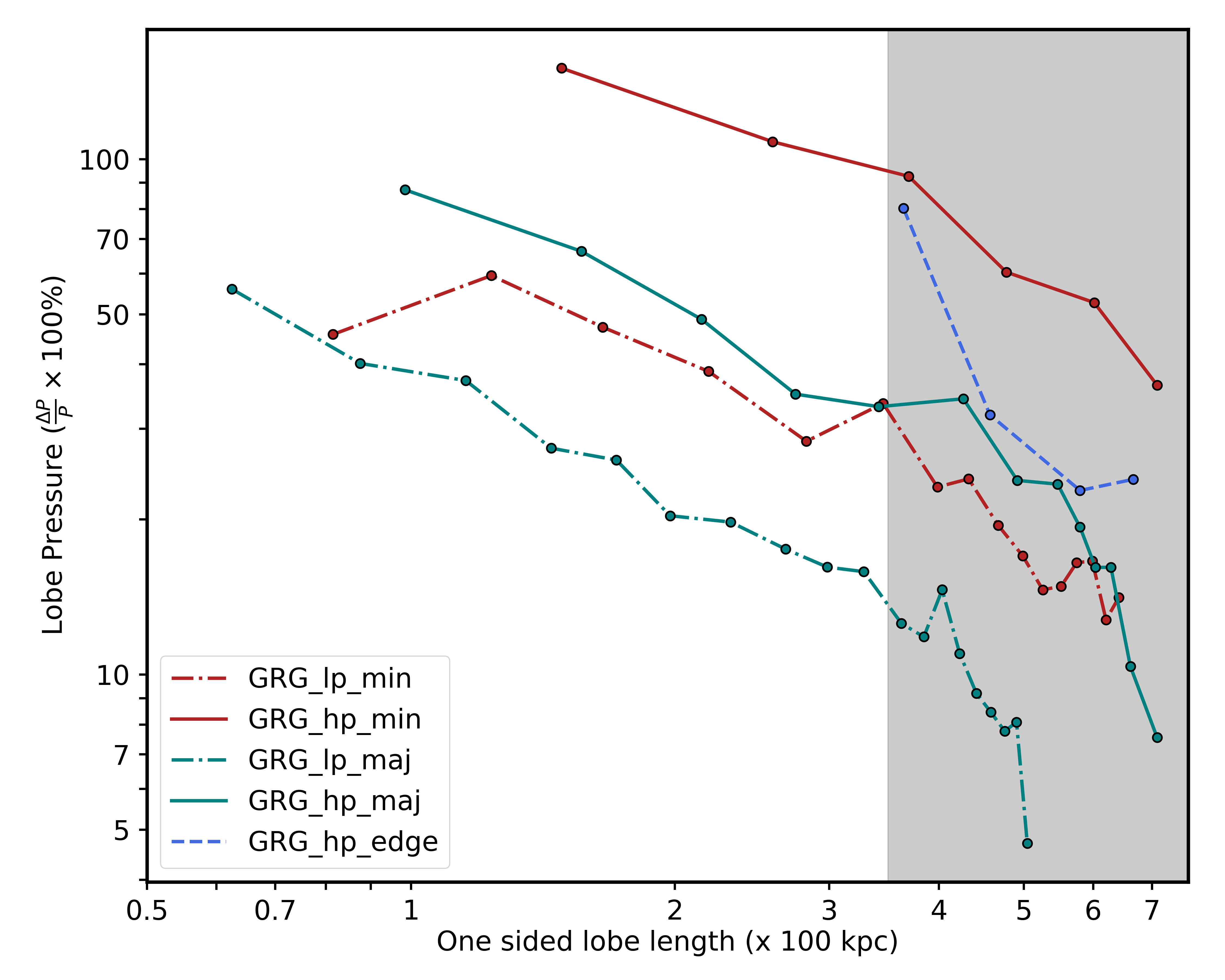}
\caption{Diagrams showing variations of lobe expansion speed (\textit{left}) and the fractional change in lobe pressure (\textit{right}) versus the one-sided lobe lengths as they evolve over time. Since the change in lobe pressure and lobe expansion speed are intricately connected variables, we plotted them alongside each other to investigate whether the giant and smaller phases of the jet exhibit any distinct or dissimilar characteristics. The shaded region extends from 350 kpc onward. For more details, refer to Section~\ref{Subsubsec:Lobe pressure and expansion speed}.}
\label{Fig:GRG_SRG} 
\end{figure*}

Since giant radio galaxies also experience smaller phases during their early evolutionary stages, it is worthwhile to explore the changes in thermodynamical quantities throughout their evolution. In our simulations, the different GRG structures evolve over varying timescales. To ensure a consistent comparison across all cases, we plotted the lobe expansion speed and the fractional change in lobe pressure against the length of the cocoon, which serves also as a proxy for temporal evolution (see, Fig.~\ref{Fig:GRG_SRG}). 

The lobe speed was plotted from 10 Myr onwards, and since the jets cover different distances by then, the initial starting points for simulation cases changes. As the jet-cocoon structures evolve further, in the case of lobe expansion speed (\textit{left} panel of Fig.~\ref{Fig:GRG_SRG}), it is found to vary marginally. However, after the one-sided jet covers a distance of about 350 kpc, the expansion speed of the lobe decreases rapidly (for four cases) or remains nearly constant (for one case). The smaller-scale fluctuations, primarily appearing at the late stages, result from a combination of the jet's ram pressure fluctuations due to its decollimation or bending and propagation along the higher pressure gradient in the outskirts of the environment.

The fractional change in pressure has been plotted starting from 20 Myr, since it represents a differential quantity (\textit{right} panel of Fig.~\ref{Fig:GRG_SRG}), defined as follows,
\begin{equation}
    \frac{\nabla P}{P} = \frac{P_i - P_{i-1}}{P_i}
\end{equation}
where $i$ is the data file number as the simulation output, saved every 9.8 Myr. As with the previous analysis, the initial length values differ across the simulation cases. The lobe pressure of all simulated structures decreases with increasing length or time, as also indicated by the fractional pressure change plot. A detailed examination of which reveals that the values of $\nabla P / P$ exhibit either marginal changes (in one case) or more rapid decreases (in four cases) after the jet reaches a length of approximately 350 kpc, compared to the earlier phases of their evolution.

These insights suggest a possible phase transition for giant radio galaxies around a particular length scale, where they transition from their smaller evolutionary stages into their giant class. This could help determine whether GRGs differ in their lobe-evolution properties compared to smaller radio galaxies. A similar phase transition behavior was observed in jet simulations by \citet{Rossi2024} while analyzing the evolution of maximum pressure in the jet-environment system. These scale-free simulations further support our findings. However, to draw definitive conclusions for GRGs, the result requires more statistical input. Future studies should explore various other parameter spaces, including the evolution of GRGs in voids \citep{Oei2022}, in super-clusters \citep{Sankhyayan2024}, and the formation of GRGs in high-powered FR IIs and low-powered FR Is to investigate GRGs in the non-preferable zones of the $\mathcal{P-D}$ diagram \citep{Delhaize2021,Simonte2024}. These derived findings, however, could guide observational studies in statistically exploring the differences between SRGs and GRGs.

\subsubsection{Comparison of lobe pressure to environmental pressure} \label{Subsubsec:Comparison of lobe pressure to environmental pressure}

To depict the pressure variation across the cocoon-ambient medium configuration more accurately, we have plotted a sliced pressure variation along the $y$-axis, perpendicular to the jet propagation, at the position of maximum cocoon width. This choice is made because the pressure variation across this direction reasonably represents the formed lobe, similar to findings by \citet{Hardcastle2013} and \citet{Malarecki2013}. 

\begin{figure*}
\centering
\includegraphics[width=2\columnwidth]{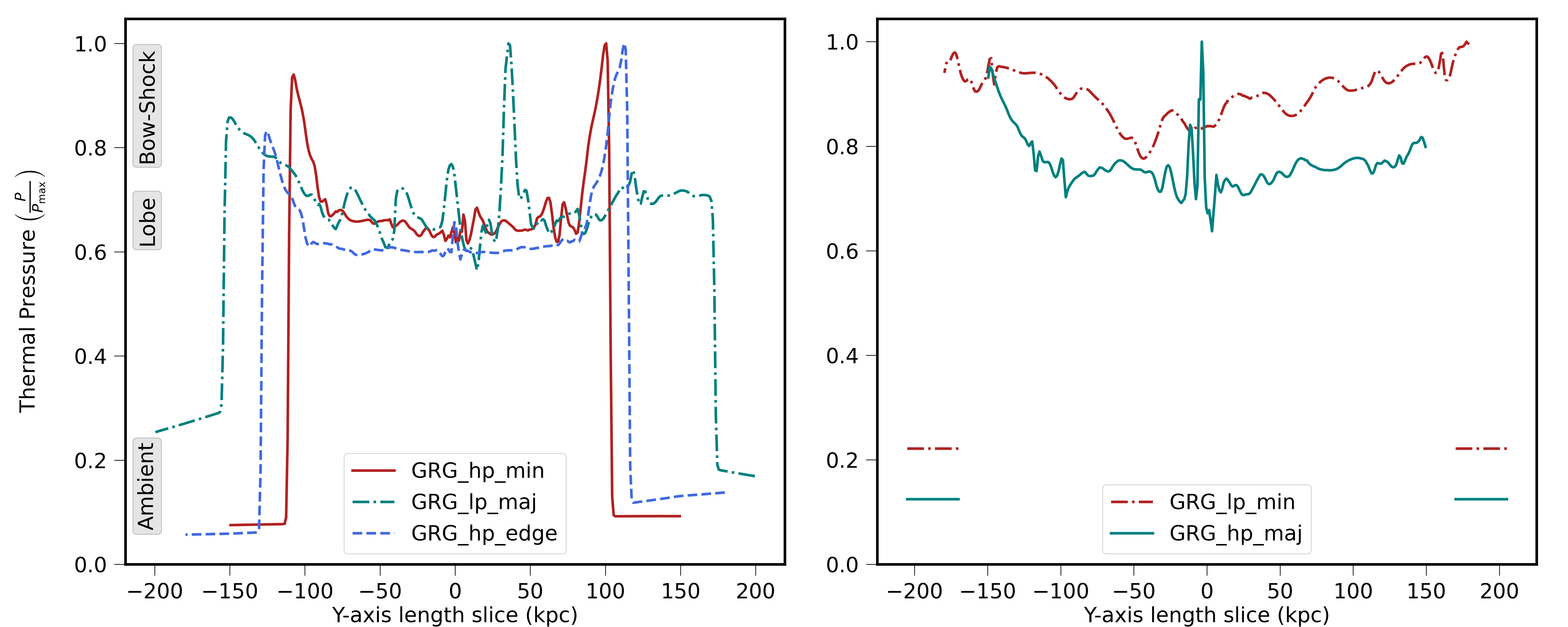}
\caption{Distribution of normalized pressure across the $y$-axis (lateral to jet propagation), indicating pressure variation within the lobe, immediate surroundings, and unperturbed ambient medium. The variation is shown at the position of maximum lobe width along the $y$-axis. The \textit{left} and \textit{right} panels represent the same situation for different runs. However, in the \textit{right} panel, the computational domain is almost entirely occupied by the lobe material in the lateral direction. So, we use the ambient medium pressure values from the unperturbed initial condition for a tentative representation.}
\label{Fig:GRG_prs_slice} 
\end{figure*}

The result is showcased in Fig.~\ref{Fig:GRG_prs_slice}, where the subplots present the same situation. However, the \textit{right} panel represents cases where the maximum width of the cocoon fully occupies the $y$-axis, and tentative relaxed ambient pressure levels are drawn with disconnected lines in the bottom as obtained from the initial conditions when the jet has not entered the domain.

Both panels of Fig.~\ref{Fig:GRG_prs_slice} illustrate a general pattern of pressure variation. The pressure within the lobe remains fairly constant, with minor fluctuations indicating the turbulent nature of the cocoon. Large spikes in pressure within the lobes are typically associated with the jet and are centered around $y = 0$, except for the green dashed line representing `GRG\_lp\_maj', where the active jet is significantly bent (see Fig.~\ref{Fig:GRG_lp_maj}). Immediately beyond the lobe, significant increase in pressure are observed, corresponding to bow shocks that effectively confine the lobe boundaries. Several features in this zone are particularly noteworthy, such as jet bending, ambient matter entrainment at the lobe interface, and buoyancy effects for sources evolving at the medium's edge, which make the bow shock strength asymmetric in both sides of the lobe. Further out, the unperturbed ambient medium exhibits lower pressure values. From the pressure jump values in Fig.~\ref{Fig:GRG_prs_slice} around the bow shock and the undisturbed ambient medium, one can estimate the Mach number (\textit{right} panel is not recommended for this estimation, as the ambient values are tentative).

It is noteworthy that the pressure difference between the bow shock and the lobe is higher for high-powered jet cases, whereas for low-powered jets, it is not as prominent. This further suggests that as the power of the jet decreases or during phases of jet activity cessation, the lobe pressure is anticipated to dissipate and gradually align with the ambient pressure \citep[see, e.g.,][]{Subrahmanyan2008,Cotton2020}. Thus, the pressure disparity between the lobe and its immediate surroundings could serve as an indicator to discern a less active or relict phase of GRGs.

\subsubsection{Evolution of magnetic field and total energy} \label{Subsubsec:Evolution of magnetic field and total energy}

Other parameters that require special attention include the evolution of the (dynamical) magnetic field and the total energy (comprising kinetic, thermal, and magnetic components).

\begin{figure*}
\centering
\includegraphics[width=2\columnwidth]{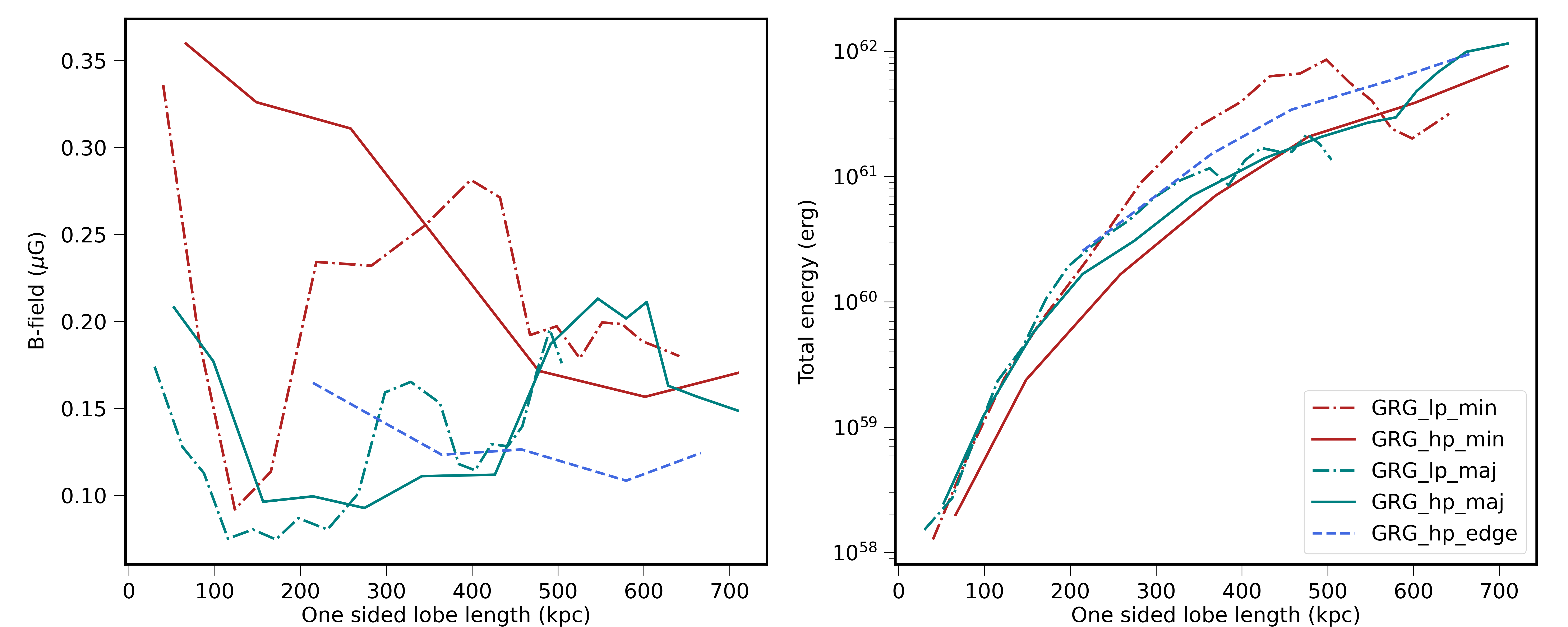}
\caption{The \textit{left} panel illustrates the length (and consequently time) evolution of the dynamical magnetic field, while the \textit{right} panel shows the evolution of total energy. Despite the seemingly random fluctuations in the magnetic field, the values tend to converge around a specific range, approximately $0.15\, \mu$G. In contrast, the total energy exhibits a relatively smooth evolution. The lines represent the five simulation cases considered in this work, all of which have a one-sided lobe length exceeding 500 kpc. Detail in Section~\ref{Subsubsec:Evolution of magnetic field and total energy}.}
\label{Fig:GRG_B_E} 
\end{figure*}

The evolution of the magnetic field (\textit{left} panel of Fig.~\ref{Fig:GRG_B_E}) does not exhibit a clear pattern throughout its evolutionary track, showing nearly random phases of growth and decay. Despite these fluctuations, the values are not distinctly different across the simulations. Regardless of the varying jet-ambient medium settings and evolutionary paths, all runs seem to converge towards a final value of around $0.15 \, \mu$G. This suggests that, regardless of the different evolution history GRGs take, if the initial conditions for the magnetic field of the jet are similar, the cocoon is expected to have a comparable final average magnetic field strength. Determining whether this pattern is common to all GRGs requires further testing with a broader parameter space and longer simulations reaching greater extents than those we have currently simulated.

We note that the estimated magnetic field here represents the dynamical magnetic field, which is expected to be lower than the equipartition field strength by a factor of $\sim 2-10$, as suggested by observational and numerical studies with varying jet power, such as those by: \citet{Croston2005,Mahatma2020,Giri2022a,Giri2022b}. Such discrepancies in magnetic field values further contribute to the inconsistency in age estimation of extended radio sources between dynamical and spectral ages. The observation that these discrepancies are found in GRGs further supports the deviation in dynamical and spectral B-field strengths \citep{Machalski2009, Machalski2011}. Based on these factors, the estimated dynamical B-field therefore appears relevant to GRGs \citep{Schoenmakers2000,Machalski2008,Sebastian2018,Andernach2021}.

Concerning the evolution of total energy (\textit{right} panel of Fig.~\ref{Fig:GRG_B_E}), the observed pattern is an expected trend relevant to extended jetted sources \citep{Giri2023}. The total energy ($E_T$) has been evaluated as,

\begin{equation} \label{eq:12}
    E_T = V_c \left[ \frac{1}{2} \rho_c v_c^2 + \frac{1}{\gamma_c - 1} P_c + \frac{B_c^2}{8\pi}\right]
\end{equation}

\noindent where the subscript $c$ is attached to parameters to represent their values associated with the cocoon determined through the tracer (tr1) values. The 1st, 2nd, and 3rd parts in Eq.~\ref{eq:12} refer to the kinetic, thermal, and magnetic components, respectively.

The total energy continuously increases as the jet injects energy into the lobe unhindered. The lobe is primarily dominated by kinetic energy, with the fraction of thermal energy being $\lesssim 0.001$ \citep{Oei2022}, and the magnetic energy being almost negligible in comparison to the kinetic energy values (fraction $\lesssim 10^{-6}$). As a result, simulations with higher jet power exhibit higher total energy than those with lower jet power. Interestingly, the low jet power cases show a slight decrease in the rate of energy increase, supporting the idea that the formation of the lobe, which deviates from self-similar expansion, allows the kinetic energy of the lobe material to relax slightly. This effect is particularly prominent in the case of 'GRG\_lp\_min'. The estimated values align with what has typically been reported for GRGs, such as by \citet{Machalski2008} and \citet{Andernach2021}.

\section{Summary} \label{Sec:Summary}
Understanding the extent to which relativistic jets can propagate has been a topic of significant interest. This paper focuses on giant radio galaxies with megaparsec-scale extents, which form relatively small subset of radio-loud AGNs. Due to their rarity and the diverse range of jet-environment configurations they are associated with, the origin of such extended structures has been a topic of considerable debate. This paper investigates the jet evolution phases of GRGs across a broad spectrum of ambient medium configurations and jet powers, aiming to numerically test the models' plausible parameter space and identify their limitations. By adjusting the ambient environment to increase (major-axis sources) and decrease (minor-axis sources) jet frustration, and varying the jet power between high and low values, we developed five distinct setups. This includes a scenario where the jet propagates along the edges of the environment with minimal resistance to the jet flow. The findings of this study can be summarised as follows, 
\begin{enumerate}
    \item We see the emergence of distinct cocoon morphologies for GRGs in varying jet-environment settings, indicating the critical role played by these two physical aspects. We demonstrated that knowledge of morphology and jet-power, when supplemented by the estimated age of the cocoon, seems to aid the understanding of their likely formation history. Although precedents for such characteristics have been found in observational studies, they are still subject to verification with current and upcoming sensitive radio observations.
    
    \item This study highlights the notably accelerated growth of the jet-cocoon structure when FR II jets advance along the peripheries of an ambient medium, extending to megaparsec scales within a timeframe comparable to smaller radio galaxies on 100s of kpc scale. This elucidates the prevalence of GRGs linked with FR II jets and their expansion in a low-density setting. This phenomenon likewise explains the development of giant radio structures on even larger scales within the standard jet propagation models.
    
    \item Regardless of the models employed, the (active) GRG lobes consistently appear over-pressured relative to the surrounding medium and are well-confined by a shocked shell of this (ambient) medium. However, no consistent pattern of over-pressure values has been observed; for example, there is no definitive correlation indicating that a high-power jet or a jet propagating along the path of maximum obstruction yields higher lobe pressures. Whereas, observing the lateral pressure variation in the lobe and its immediate surroundings, depending on whether the pressure decrease is sharp or gradual, may help distinguish whether a GRG is in a less active or relic phase.

    \item Measuring the axial ratio (ratio of length to width) seems promising for understanding whether GRGs adhere to self-similar expansion patterns. While high-powered jets generally exhibit such expansion, low-powered jets diverge from this behavior primarily because of lobe formation, which significantly decelerates lobe expansion. However, this assessment may be influenced by the projection effect, a factor we aim to validate in a separate study.

    \item While comparing giant radio galaxies with their smaller phases, a shift in the variation of lobe expansion speed and fractional change in lobe pressure has been observed around a length scale of $\sim 350$ kpc (one-sided) for our formulated simulation settings, indicating the likely presence of a transition between SRGs and GRGs. However, this claim needs to be rigorously tested with even more extended parameter sets, supplemented by observational input (discussed below) to confirm its general appearance.

    \item The (dynamical) magnetic field within the cocoon of giant radio galaxies exhibits no discernible pattern in its temporal variation. However, regardless of the models employed, consistent convergence to similar magnetic field strengths occurs when the initial magnetic field strength of the launched jet remains the same for all cases. The lobes of GRGs typically remain kinetically dominated.

    \item The formation of off-axis lobes in GRGs resembling wings is a distinct feature that arises due to the influence of the tri-axial dynamics of large-scale medium and grows alongside the jet's giant phases. Detection of these passively evolving lobes by present-day and forthcoming sensitive radio telescopes holds the potential to shed light on whether winged sources constitute a minor subset of radio-loud AGNs or if the prevalence of winged sources among radio galaxies is limited by the sensitivity thresholds of current radio telescopes.
\end{enumerate}

This study focuses on the dynamical behavior of GRGs, emphasizing the morphology formed and variations in magneto-hydrodynamical properties based on a set of jet-environment configurations. While shedding light on GRG growth, a further extension of this work would involve investigating the micro-physical processes occurring inside the jet cocoon, such as particle reacceleration or cooling effects \citep{Vaidya2018,Mukherjee2021,Kundu2022}, which are believed to have significant impacts on GRG appearances, particularly in high-redshifts \citep{Hintzen1983,Schoenmakers2000,Konar2004}. It is also crucial to extend numerical exploration of GRG properties such as those found in voids \citep{Oei2022} or in super-clusters \citep{Sankhyayan2024}, in order to assess the general applicability of the highlighted models. Moreover, examining the formation of GRGs in high-powered FR IIs and low-powered FR Is is essential to understand their scarcity in less favored regions of the $\mathcal{P-D}$ diagram \citep{Delhaize2021,Simonte2024}. We also underscore the necessity of statistically modeling these giant jetted systems within expansive cosmological contexts \citep[e.g.,][]{Dave2019,Nelson2019} to understand their feedback effects on the environment and their potential utility as probes for characterizing the Warm-Hot Intergalactic Medium.

Expanding upon this discussion, it is worth noting that a group of parameters in our simulated models, although carefully chosen to initially mimic standard jet-environment systems, are held constant at single values. These include, for example, the density contrast between the jet and ambient medium, the configuration and strength of the jet magnetic field, the eccentricity of the environment and parameters related to the King’s $\beta$-profile. Previous studies on various astrophysical jet systems have underscored their critical role in stabilizing and sustaining a jet’s evolutionary path. These studies demonstrated how initial magnetic field strength and configuration can influence the development of Kelvin-Helmholtz or kink instabilities, ultimately affecting entrainment, jet expansion speed, and long-term evolution \citep{Mukherjee2020, Rossi2024}. Work by \citet{Rossi2017} has shown that different jet-to-ambient density contrast values impact jet flow speed, which, in turn, affects the linear growth of a jet system over time \citep[see also its effect on FR-I type jets;][]{Rossi2008}. Focusing on the role of environmental eccentricity, \citet{Hodges-Kluck2011} emphasized its impact on the development of prominent wings in X-shaped systems, facilitating the formation of notable XRGs (with a high wing-to-lobe length ratio) as eccentricity increases. A discussion on the influence of environmental parameters on the evolution of powerful jets can also be found in, for example, \citet{Hardcastle2018}. 

It can thus be speculated that understanding the development of giant phases in straight jets is complex, especially from a theoretical perspective, given the combined influence of multiple underlying mechanisms. Consequently, more focused studies on these giant RGs are essential, and some of these will be addressed in our forthcoming studies. 

\begin{acknowledgements}
GG is a postdoctoral fellow under the sponsorship of the South African Radio Astronomy Observatory (SARAO). The financial assistance of the SARAO towards this research is hereby acknowledged (\url{https://www.sarao.ac.za/}). GG also acknowledges the pivotal discussions with William D. Cotton at the MeerKAT$@$5 conference, which sparked this study. JB acknowledges the support from Department of Physics and Electronics, Christ University, Bangalore. RPD acknowledges funding by the South African Research Chairs Initiative of the Department of Science and Innovation and National Research Foundation (Grant ID: 77948). The authors acknowledge the Centre for High Performance Computing (CHPC), South Africa, for providing computational resources to this research project (\url{https://www.chpc.ac.za/}). We acknowledge the use of the ilifu cloud computing facility - \url{https://www.ilifu.ac.za/}, a partnership between the University of Cape Town, the University of the Western Cape, Stellenbosch University, Sol Plaatje University, the Cape Peninsula University of Technology and the South African Radio Astronomy Observatory. The ilifu facility is supported by contributions from the Inter-University Institute for Data Intensive Astronomy (IDIA - a partnership between the University of Cape Town, the University of Pretoria and the University of the Western Cape), the Computational Biology division at UCT and the Data Intensive Research Initiative of South Africa (DIRISA). We extend our gratitude to Jeremy Smith (Inter-University Institute for Data Intensive Astronomy) and Anna Bosman (University of Pretoria) for their assistance in this regard. GG acknowledges financial support from the IDIA for a research visit to the University of Cape Town.
\end{acknowledgements}

%
%


\bibliographystyle{aa} 
\bibliography{sample1} 

\begin{appendix}
\section{Effects of Resolution Enhancement} \label{AppendixA: Effects of Resolution Enhancement}
\subsection{Setting up the resolution runs}
In light of the lack of numerical studies available to validate models explaining the growth of giant radio galaxies, this work has been developed within a fully 3D relativistic MHD framework in realistic jet-environment settings, representative of typical GRG formation conditions. Furthermore, symmetry-breaking operations inherent to 3D dynamics have been incorporated, lending this study a distinctive methodological approach. 

However, as with all simulation models, it is essential to assess the effect of resolution on the results, as achieving convergence in terms of resolution in 3D is often computationally challenging. For instance, there is general consensus that simulating jetted sources at high resolution requires approximately 8 grid zones within the jet radius to capture relevant details accurately \citep{Stone1994,Aloy1999}. However, studies such as \citet{Mignone2010}, \citet{Hodges-Kluck2011}, and \citet{Giri2022b} have demonstrated the use of even higher grid resolution within the jet radius ($\sim 15$ grid cells) to better capture transverse structures, such as shock formation along the jet beam and the influence of pressure gradients on evolving structures. A recent study by \citet{Dubey2023} employed an even greater number of grid cells within the jet radius ($\sim 25$), noting that while the overall external morphology may appear similar with lower resolution, key micro-physical processes within the jet-cocoon, may vary significantly. 

Achieving the grid resolution necessary to capture the majority of underlying physics is ideal but challenging, particularly for simulating jets at giant scales with the computing resources currently available. For instance, using 20 grid cells per jet radius would require a 3D grid of $[14400 \times 8000 \times 8000]$ zones for a single run, which poses significant computational demands. Thereby, the highlighted earlier studies have either omitted several physical aspects to focus on a broader overview or have limited their simulations to the jet’s evolution at smaller scales. Driven by the goal of resolving both the jet and the large-scale lobe structures of giant radio galaxies with equal importance, and with our computational resources distributed across five models, we achieved only 2 grid cells per jet diameter. Consequently, a comparison with a slightly higher-resolution run has become essential. 

We conducted our experiment on the case `GRG\_hp\_min', considering it a powerful source that is more susceptible to instabilities and capable of generating more shocks within the cocoon. The model presented in Table~\ref{Tab:Parametric_space} for this case is designated as `Reference\_Res'. We then conducted two additional sets of simulations: one with 4 grid zones per jet diameter, resulting in a grid of $[1440 \times 600 \times 600]$ to cover a domain of $[720 \times 300 \times 300]$ kpc$^3$, and the other performed in a zoom-in manner focusing on the early phases of jet evolution, with a grid of $[1050 \times 700 \times 700]$ to cover a domain of $[210 \times 140 \times 140]$ kpc$^3$, raising 10 grid cells per jet diameter. These simulations were designated as `High\_Res (extended)' and `High\_Res (zoom-in)', respectively. A brief overview of the simulations are presented in Table~\ref{Tab:Resolution_Test}.

\begin{table*}
\caption{Different resolution runs for the case: `GRG\_hp\_min'.}
\begin{center}
\begin{tabular}{ |l|c|c|c|c|l| } 
 \hline
 Simulation& Domain & Number of & Cells per & Simulation & Remarks  \\
 Label & (kpc$^3$) & Grids & Jet diameter & Time (Myr) &  \\
 \hline
 Reference\_Res & $720\, \times 300 \, \times 300 $ & $720\, \times 300 \, \times 300 $ & 2 & 68.5  & Setup reported in Table~\ref{Tab:Parametric_space},\\
 &&&& ($T_L$) & Jet reached domain length at $T_L$\\
 \hline
 High\_Res (extended) & $720\, \times 300 \, \times 300 $ & $1440\, \times 600 \, \times 600 $ & 4 & 43.1 & Higher resolution, Full Domain, \\
 &&&&& Followed to $\sim 2/3$rd of $T_L$\\
 \hline
 High\_Res (zoom-in) & $210\, \times 140 \, \times 140 $ & $1050\, \times 700 \, \times 700 $ & 10 & 6.5 & Followed the initial jet evolution\\
 &&&&& phases in a smaller domain\\
  \hline
\end{tabular}

\label{Tab:Resolution_Test}
\end{center}
\small
\textbf{Notes.} Beginning with column 1, where three simulations are labeled, this table provides an overview of the resolution runs: columns 2 and 3 present the domain size and grid count, column 4 indicates the number of grid cells used to resolve the initial jet flow, column 5 shows the simulation duration, and column 6 provides concise details about each run.
\end{table*}

\subsection{Insights from the resolution runs}
\begin{figure}
\centering
\includegraphics[width=\columnwidth]{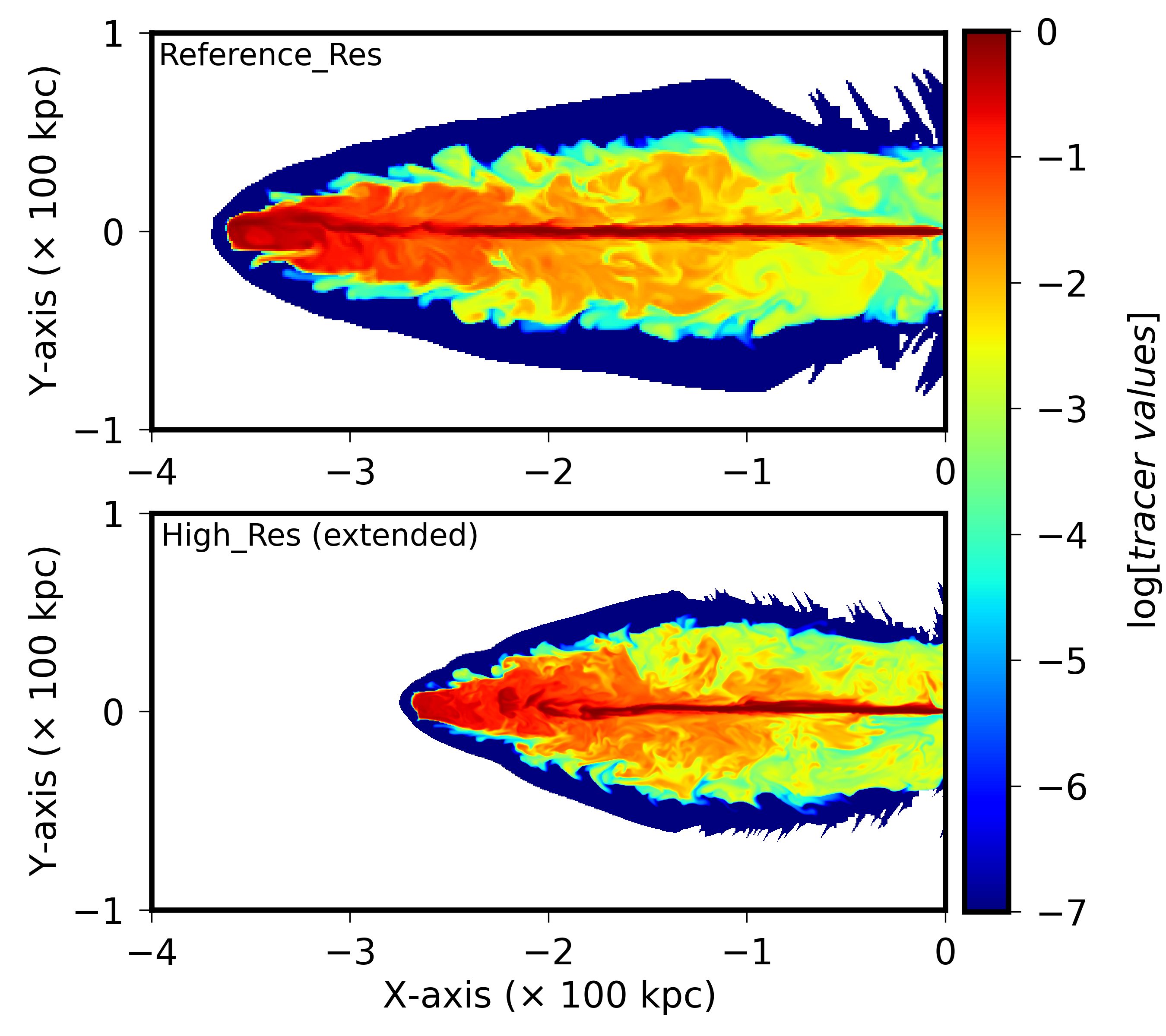}
\caption{Comparison of $x-y\, (z = 0)$ jet structures (based on tracer values) obtained from the `Reference\_Res' and `High\_Res (extended)' runs at $\sim 39$ Myr (dynamical age). These two cases follow the longer-term evolution of the jet.}
\label{Fig:GRG_Res_Img} 
\end{figure}

From Fig.~\ref{Fig:GRG_Res_Img}, the immediate picture that arises is the lesser expansion rate of the jet-cocoon structure in the higher-resolution case compared to the reference one. This result is not unexpected, given that higher-resolution runs can better resolve the effect of pressure gradient forces in the ambient medium (responsible for lateral expansion of the cocoon); additionally, the jet beam is now susceptible to instabilities, which affect the jet expansion speed. Regardless, it is clear that a high-powered jet propagating along the minor axis of the environment (i.e., in a direction of lower jet frustration) undergoes rapid lengthwise growth ($\sim 300$ kpc of one-sided jet-length covered in $39$ Myr of dynamical age). Consequently, as noted earlier, such a jet would not require much time to evolve into a standard giant radio galaxy.

\begin{figure}
\centering
\includegraphics[width=\columnwidth]{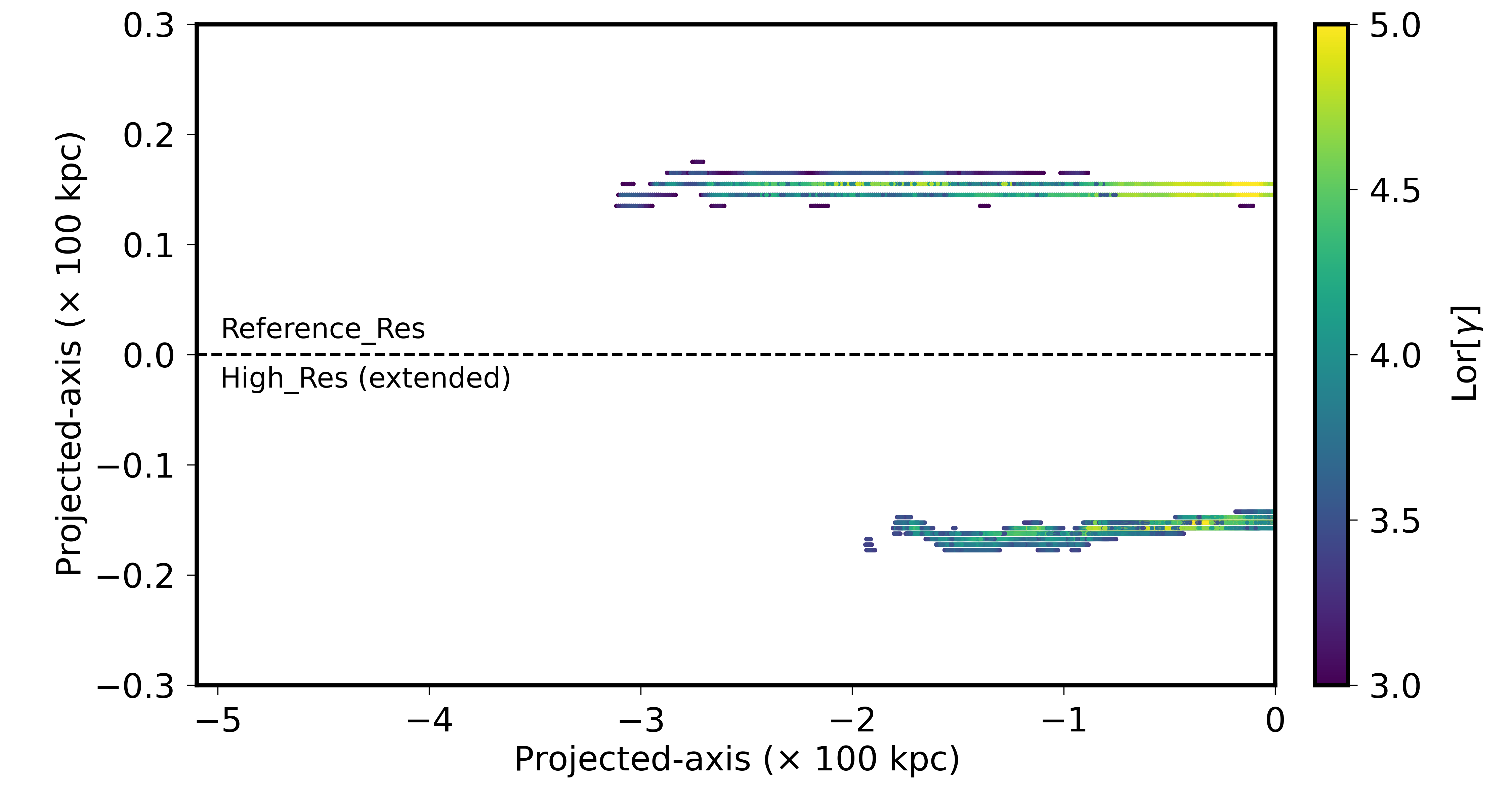}
\caption{Projected visualization of the 3D jet-beam structures for the `Reference\_Res' and `High\_Res (extended)' cases, represented by Lorentz factor values (highest few thousand values) shown in the colorbar. The jet positions are shifted vertically to accommodate both cases within a single plot. Notably, the bending of the jet beam in the higher-resolution case is more pronounced than in the reference case, suggesting the development of instabilities with the jet's motion.}
\label{Fig:GRG_LorG_Img} 
\end{figure}

To illustrate the emergence of instabilities in the jet beam, which likely impact the jet expansion speed, we plotted a projected map of the highest 2000 Lorentz factor values for the `Reference\_Res' case and 4000 values for the `High\_Res (extended)' case (as it has twice the resolution). These points, expected to align with the primary jet beam, are projected onto a plane (mimics a typical 3D configuration) where Lorentz factor values are represented through a color bar, formulating a 2D histogram. This visualization is shown in Fig.~\ref{Fig:GRG_LorG_Img}, where the two jet beams are slightly offset to fit within a single plot. The bending of the jet is significantly more evident in the higher-resolution case than in the reference case, where it is only mildly observed. This highlights that jet instabilities linked to powerful jets in giant radio galaxies are a natural occurrence that can impact the linear motion of the jets. Given the importance of these instabilities, a dedicated study is necessary to understand the growth history of these micro-features, as their evolution is also influenced by multiple factors, including the configuration and strength of the magnetic field, the density contrast between the jet and its surrounding medium. Recent high-resolution observational studies have also documented instabilities of this nature, reinforcing the relevance of these findings in understanding jet behavior in giant radio galaxies \citep{Dabhade2022}.

\begin{figure*}
\centering
\includegraphics[width=2\columnwidth]{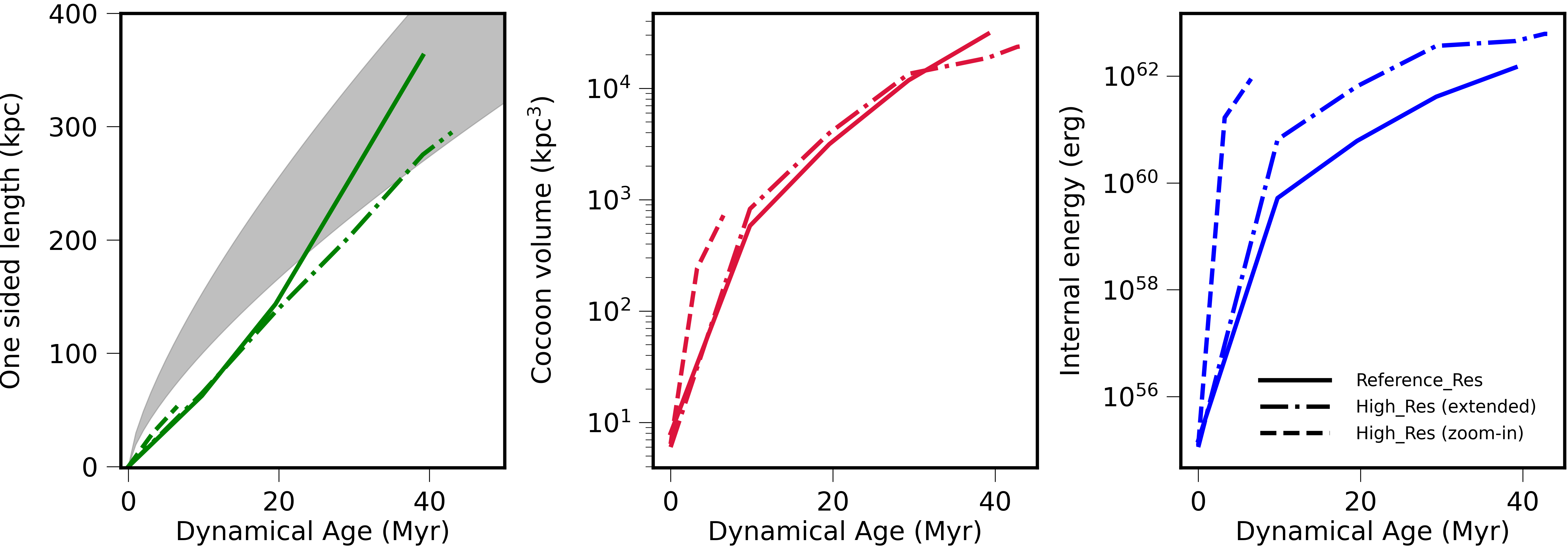}
\caption{Evolution patterns of the internal and external behavior of the simulated jet-cocoon system over time for the three resolution cases discussed in Table~\ref{Tab:Resolution_Test}. The parameters plotted here are the lengthwise cocoon growth (\textit{left}), cocoon volume growth (\textit{middle}), and internal energy growth (\textit{right}) across the three systems simulated at varying number of grid cells inside jet diameter.}
\label{Fig:GRG_thermo_properties} 
\end{figure*}

Moving forward, we aimed to investigate the impact of resolution on external properties (jet length and cocoon volume) and internal property (internal energy) across the three runs, including the `High\_Res (zoom-in)' case. First, we examined the evolution of jet length with dynamical age for the three cases, as illustrated in Fig.~\ref{Fig:GRG_thermo_properties} (\textit{left}). Notably, the initial development of the jet structure around the denser inner core is consistent across all three cases. However, as time progresses, instabilities in the jet beam emerge, resulting in differences in jets' linear evolution with age. Despite these variations, the evolution ultimately aligns with the theoretically permitted zones, suggesting that the results obtained at the reference resolution do not introduce any significant deviations from the established conclusions. This is further supported by the observation that the evolution of cocoon volume for all three cases closely follows one another over time. The deviations are not markedly different; in fact, the values eventually align with the primary evolution pattern (Fig.~\ref{Fig:GRG_thermo_properties} (\textit{middle})). This indicates that the 3D morphological characteristics, in terms of topology, remain consistent among the cases. In the case of internal energy variation, differences in values are observed (the patterns of evolution remain consistent though; Fig.~\ref{Fig:GRG_thermo_properties} (\textit{right})). As anticipated, the highest resolution run, which focuses on small-scale jet evolution, yields elevated energy values due to its ability to resolve more shock structures generated by the jet flow, resulting in increased heating of the cocoon material. In contrast, the reference resolution case exhibits lower heating, as expected from this perspective. However, if we consider the `High\_Res (extended)' case, it ultimately converges with the `Reference\_Res' case over a longer evolution timeframe. Thus, it may be premature to assert that increasing resolution will consistently lead to higher internal energy detection within the cocoon of simulated giant radio galaxies. As the system ages, shock strength likely diminishes, referring that the observed internal energy increase could be an initial feature that subsides over time. Additionally, the cocoon volume evolution shows similar converging behavior at later stages, indicating that any early energy enhancement may smooth out as the system progresses.

\section{Observability of Recollimation Shocks} \label{Appendix:Observability of Recollimation Shocks}
\begin{figure*}
\centering
\includegraphics[width=2\columnwidth]{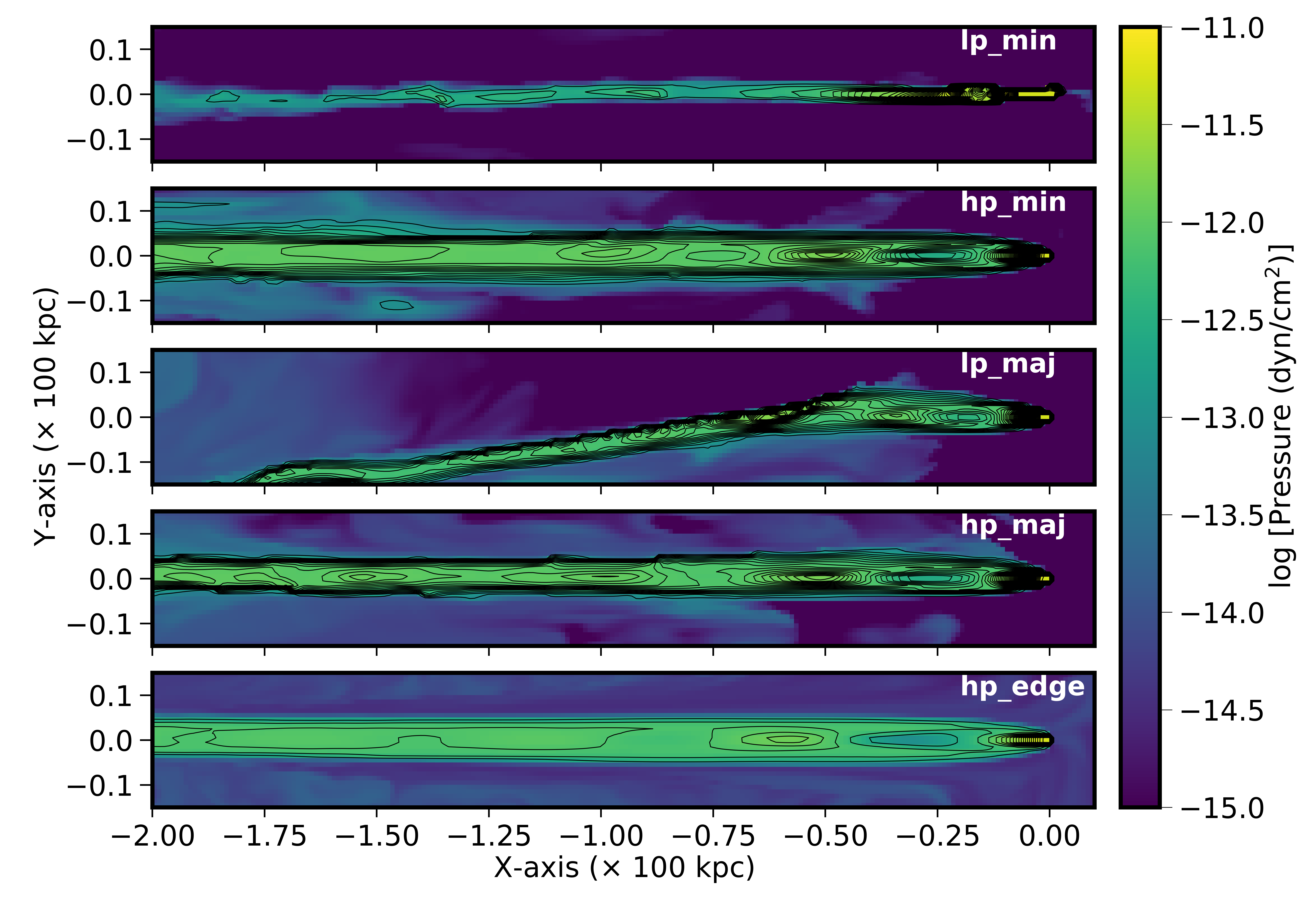}
\caption{2D slices of pressure values (indicated by the colorbar) are plotted in the $x-y\, (z = 0)$ plane, zoomed into the jet-beam region, and show the developed structures of the five primary simulations from Table~\ref{Tab:Parametric_space} at their respective evolutionary stages, as seen in Fig.~\ref{Fig:GRG_lp_min} - \ref{Fig:GRG_hp_edge} (with each case labeled in the subplots). Pressure contours in black solid-lines have been overlaid to illustrate the complex pressure distribution within the jet beam, highlighting zones of compression and rarefaction. This approach helps visualize the formation of recollimation shocks, particularly prominent in high-powered jet cases. In lower-powered jets, these shocks are noticeable in early evolution length-scales, though compression zones continue to appear as the jets undergo bending. This analysis also reveals that the jet-beam expands across multiple grid zones shortly after injection, effectively reducing the impact of resolution constraints.}
\label{Fig:GRG_prs_shock} 
\end{figure*}

Our main simulations (Table~\ref{Tab:Parametric_space}) began with only 2 grid cells across the jet diameter. However, due to the initial jet pressure being $\sim 1.4$ to $3.1$ times higher than that of the surrounding environment near the injection zone, the jet rapidly expands to occupy multiple grid cells as propagation initiates. This expansion is evident in Fig.~\ref{Fig:GRG_LorG_Img}, where the Lorentz factor values at the launching site quickly spread across several grid cells. Additional hints of this expansion appears in the 3D volume-rendered images in Fig.~\ref{Fig:GRG_lp_min} - \ref{Fig:GRG_hp_edge}. However, to investigate this matter in detail, we have plotted the pressure values within the jet structures for our main runs at specific stages of their development (ages as mentioned in Fig.~\ref{Fig:GRG_lp_min} - \ref{Fig:GRG_hp_edge}). Each plot shows a 2D zoom-in slice of the developed jet structure in the $x-y \, (z = 0)$ plane, as illustrated in the Fig.~\ref{Fig:GRG_prs_shock}. The figure clearly illustrates the expansion across all cases, likely diminishing the impact of resolution limitations as the expanded jet occupies multiple grid cells. A comprehensive analysis of these patterns is still pending, but the influence of MHD parameters on this behavior is evident. Our planned parameter studies will explore these effects in greater depth in future work.

We further examined small-scale pressure behavior within the jet-beam by overlaying pressure contours on the pressure map. This approach highlights zones of compression and rarefaction within the jet-beam structure for each case. As shown in Fig.~\ref{Fig:GRG_prs_shock}, high-powered jets display distinct zones of pressure compression, with oscillatory variations in the Lorentz factor (not shown here) suggesting the formation of recollimation shocks. Hints of this oscillatory behavior can also be observed in the 3D volume-rendered plots of density presented in Figs.~\ref{Fig:GRG_hp_min}, \ref{Fig:GRG_hp_maj}, and \ref{Fig:GRG_hp_edge}. In contrast, low-powered jets exhibit pressure compression close to the launching sites, which gradually diminishes over distance. However, as low-powered jets are more susceptible to instabilities under the same MHD parameters compared to high-powered jets, bending occurs further along their length, leading to renewed compression zones at these bend locations \citep[e.g.,][]{Condon2021,Wezgowiec2024}. Testing the visibility of these compressed zones in sky-projected emission maps is an area we plan to explore in future studies.

\end{appendix}

\end{document}